\documentclass[a4paper,12pt]{article}
\usepackage{jcappub} 
\usepackage{cite}
\usepackage{graphicx}
\usepackage{enumerate}
\usepackage{hyperref}
\usepackage{cancel}
\usepackage{color}
\usepackage{graphicx}
\usepackage{amsmath}
\usepackage{amsfonts}
\usepackage{amssymb}
\usepackage{rotating}
\usepackage{booktabs}
\usepackage{xcolor}
\usepackage{soul}

\def\red#1{\textcolor{red}{#1}}

\def\comment#1{}

\title{\boldmath 
Massive particle pair production and oscillation in Friedman Universe:
its effect on inflation
}

\author{She-Sheng Xue}

\affiliation{ICRANet Piazzale della Repubblica, 10 -65122, Pescara, Italy, \\ Physics Department, University of Rome La Sapienza, \\ P.le Aldo Moro 5, I–00185 Rome, Italy\\
INFN, Sezione di Perugia, Via A. Pascoli, I-06123, Perugia, Italy
}

\emailAdd{xue@icra.it and shesheng.xue@gmail.com} 

\abstract{We study the classical Friedman equations for the time-varying cosmological term $\tilde\Lambda$ 
and Hubble function $H$, together with quantised field equations for the production of massive $M\gg H$ particles, namely, the $\tilde\Lambda$CDM scenario of dark energy and matter interactions. Classical slow components ${\mathcal O}(H^{-1})$ are separated from quantum fast components ${\mathcal O}(M^{-1})$. The former obeys the Friedman equations, and the latter obeys a set of nonlinear differential equations. Numerically solving equations 
for quantum fast components, we find the production and oscillation of massive 
particle-antiparticle pairs in microscopic time scale ${\mathcal O}(M^{-1})$. Their density and pressure averages over microscopic time do not vanish. 
It implies the formation of a massive pair plasma state in macroscopic time scale ${\mathcal O}(H^{-1})$, whose effective density and pressure contribute to the Friedman equations. Considering the inflation driven by the time-varying cosmological term and slowed down by the massive pair plasma state, we obtain the relation of spectral index and tensor-to-scalar ratio in agreement with recent observations. We discuss the singularity-free
pre-inflation, the CMB large-scale anomaly, and dark-matter density perturbations imprinting on power spectra.
}

\begin{document}
\maketitle
\flushbottom

%
\section{\bf Introduction}\label{int}
The gravitational particle creation in Friedman Universe expansion is an 
important theoretical issue \cite{PhysRevLett.21.562, PhysRev.183.1057,  PhysRevD.3.346, Zeldovich1971, Birrell1984} that has been intensively studied for decades \cite{Mamaev1976, Zeldovich1977, Starobinsky:1979ty, Mottola1985, Habib2000, Anderson2014, Anderson2014a, Landete2014}. Based on adiabaticity and non-back-reaction approximation for a slowly time-varying Hubble function $H(t)$, one adopted the semi-classical WKB approaches to calculating 
the particle production rate, which is exponentially suppressed $e^{-M/H}$ for massive particles $M\gg H$. However, the non-adiabatic back-reactions of  
particle creations on the Hubble function can be large and have to be taken into account.
The non-adiabatic back-reactions of massive particle productions have a quantum time scale ${\mathcal O}(1/M)$ that is much smaller than classical Universe evolution time 
scale ${\mathcal O}(1/H)$. To properly include the back-reaction of particle production on Universe evolution, one should separate fast components ${\mathcal O}(1/M)$ from slow components ${\mathcal O}(1/H)$ in the Friedman equation. 
Many efforts \cite{Parker1973,Starobinsky1980,Starobinsky1982,Ford1987,Kolb1996,Greene1997,Kolb1998,Chung1998,Chung2001,Chung2000,Chung2003,Chung2005,Ema2016,Chung2019,Ema2018,Li2019,Xue2019,Xue2020} have been made to study non-adiabatic 
back-reaction and understand massive particle productions without exponential suppression. It is important for reheating, possibly accounting for massive dark matter and total entropy of the present Universe \cite{Starobinsky1980,Starobinsky1982,Kofman1994,Kofman1997,Kuzmin1998,Kuzmin1999,Kolb1999,Bertone2005,Bassett2006,Kolb2007,Allahverdi2010,Frolov2010,Hall2010,Feng2010,Gorbunov2012,Amin2014,Ema2015,Garny2016,Garny2018,Kolb2017,Hashiba2019,Hashiba2019a,Haro2019,Xue2020a}. 

In this article, we consider the $\tilde\Lambda$CDM scenario \cite{Xue2015}
of dark energy and matter interactions, in which the cosmological term $\tilde\Lambda$ is time-varying
and the Friedman equations for a flat Universe become
\begin{eqnarray}
H^2=\frac{8\pi G}{3}\rho;\quad \dot H=-\frac{8\pi G}{2}(\rho + p)=-\frac{8\pi G}{2}(\rho_{_M}+ p_{_M}),\label{friedman}
\end{eqnarray}
where energy density $\rho\equiv \rho_{_M}+\rho_{_\Lambda}$ and pressure 
$p\equiv p_{_M}+p_{_\Lambda}$. 
Equation of state $p_{_\Lambda}=-\rho_{_\Lambda}$ is for the cosmological constant term (dark energy).  Equation of state $p_{_M}=\omega_{_M}\rho_{_M}$ is 
for the matter that represents relativistic (radiation) and non-relativistic components.
The second Equation of (\ref{friedman}) is the generalised 
conservation law (Bianchi identity) for including time-varying cosmological term
$\rho_{_\Lambda}(t)\equiv \tilde\Lambda/(8\pi G)$. It reduces to the usual Equation 
$\dot \rho_{_M} + (1+\omega_{_M})H\rho_{_M}=0$ for 
time-constant $\rho_{_\Lambda}$.  
The second Equation of (\ref{friedman}) shows that due to the matter's 
gravitational attractive nature, $\dot H <0$ and $H$ decreases in time. 
Equations (\ref{friedman}) are not the same as the Friedman equations with the constant cosmological term $\Lambda$ or the scalar field $\phi$ of inflation 
potential $V(\phi)$.

\section{Slow and fast components' separation}

In the $\tilde\Lambda$CDM scenario, we adopt the 
approach \cite{Chung2019} to describe the decomposition of slow and fast components: scale factor $a=a_{\rm slow}+a_{\rm fast}$, Hubble function $H=H_{\rm slow}+H_{\rm fast}$, cosmological term and matter density $\rho_{_{\Lambda,M}}=\rho^{\rm slow}_{_{\Lambda,M}}+\rho^{\rm fast}_{_{\Lambda,M}}$ and pressure $p_{_{\Lambda,M}}=p^{\rm slow}_{_{\Lambda,M}}+p^{\rm fast}_{_{\Lambda,M}}$. The fast components vary much faster in time, but their amplitudes are much smaller than the slow components. 
According to the order of small ratio $\lambda$ of fast and slow components, the Friedman equations (\ref{friedman}) are decomposed into two sets. The slow components 
${\mathcal O}(\lambda^0)$
obey the same equations as usual Friedman equations (``macroscopic'' ${\mathcal O}(H_{\rm slow}^{-1})$ equations)
\begin{eqnarray}
H_{\rm slow}^2&=&\frac{8\pi G}{3}(\rho_{_M}^{\rm slow}+\rho_{_\Lambda}^{\rm slow});\nonumber\\
\quad \dot H_{\rm slow}&=&-\frac{8\pi G}{2}(\rho_{_M}^{\rm slow} +p_{_M}^{\rm slow}),
\label{sfriedman}
\end{eqnarray}
where $H_{\rm slow}=\dot a_{\rm slow}/a\approx 
\dot a_{\rm slow}/a_{\rm slow}$, 
time derivatives $\dot H_{\rm slow}$ and $\dot a_{\rm slow}$ 
relate to the macroscopic 
``slow'' time variation scale ${\mathcal O}(1/H)$.
The faster components ${\mathcal O}(\lambda^1)$ obey new (``microscopic'' ${\mathcal O}(M^{-1})$ equations),
\begin{eqnarray}
H_{\rm fast}&=&\frac{8\pi G}{2\times 3H_{\rm slow}}(\rho_{_M}^{\rm fast}+\rho_{_\Lambda}^{\rm fast});\nonumber\\ 
\dot H_{\rm fast}&=&-\frac{8\pi G}{2}(\rho_{_M}^{\rm fast} +p_{_M}^{\rm fast}),
\label{ffriedman}
\end{eqnarray} 
where 
$H_{\rm fast}=\dot a_{\rm fast}/a\approx \dot a_{\rm fast}/a_{\rm slow}$, 
time derivatives $\dot H_{\rm fast}$ and 
$\dot a_{\rm fast}$ relate to the microscopic ``fast'' time variation scale ${\mathcal O}(1/M)$, and slow components are approximated as constants in ``fast'' time variation. For the cosmological term,
equation of state $p_{_\Lambda}=-\rho_{_\Lambda}$ becomes 
$p_{_\Lambda}^{\rm slow,fast}=-\rho_{_\Lambda}^{\rm slow,fast}$ respectively at order ${\mathcal O}(\lambda^0)$ 
and ${\mathcal O}(\lambda^1)$. In due course, we shall clarify
the equation of state $p_{_M}=\omega_{_M}\rho_{_M}$ for the matter term, associating to the fast and slow components respectively. 
Equations (\ref{ffriedman}) for the fast components are different from their counterparts \cite{Chung2019} for the case of Friedman equations 
with a single inflation field $\phi$ and its potential $V(\phi)$. They are novel equations to investigate the nature of dark energy and matter interactions in the 
$\tilde\Lambda$CDM scenario. 

In the fast component Equation (\ref{ffriedman}), we adopt the approach \cite{Parker1973} to describe the fast components of matter density $\rho_{_M}^{\rm fast}$ and pressure $p_{_M}^{\rm fast}$. They are due to the non-adiabatic production of particle and antiparticle pairs in fast time variation $H_{\rm fast}=\dot a_{\rm fast}/ a_{\rm slow}$ and its time derivative $\dot H_{\rm fast}$.
As new results, we find quantum coherent oscillation of fast and microscopic components $H_{\rm fast}$, $\rho_{_\Lambda}^{\rm fast}$, $\rho_{_M}^{\rm fast}$ and $p_{_M}^{\rm fast}$, due to microscopic back reactions at the time scale ${\mathcal O}(M^{-1})$. The quantum pair production and oscillation of $\rho_{_M}^{\rm fast}$ and $p_{_M}^{\rm fast}$ can form a macroscopic state of massive pair plasma, contributing to slow and macroscopic components $\rho_{_M}^{\rm slow}$ and $p_{_M}^{\rm slow}$ at the time scale ${\mathcal O}(H^{-1})$. In the $\tilde\Lambda$CDM scenario, we consider the time-varying cosmological term $\rho_{_\Lambda}^{\rm slow}$ drives the inflation (quasi-de Sitter phase) \cite{Starobinsky:1979ty, Starobinsky1980, Linde:1981mu, Mukhanov:1982nu, Kallosh:2021mnu}.  
We study how such a macroscopic state of massive pair plasma affects (back-reacts on) the Friedman equation (\ref{sfriedman}) by slowly decreasing $\rho_{_\Lambda}^{\rm slow}$ and $H_{\rm slow}$, leading to slowing-down effects on the inflation. 

\section{Quantum pair production and oscillation}\label{qppo}
A quantised massive scalar matter field inside the Hubble sphere volume 
$V\sim H^{-3}_{\rm slow}$ of Friedman Universe reads
\begin{eqnarray}
\Phi({\bf x},t)&=&\sum_n A_n Y_n({\bf x})\psi_n(t).
\label{qfield}
\end{eqnarray} 
Here we assume that the field exponentially vanishes outside the horizon $H^{-1}_{\rm slow}$, i.e., the particle horizon $(a_{\rm slow}H_{\rm slow})^{-1}$ of comoving Hubble radius,
and 
$\int_V Y_n({\bf x}) Y^\dagger_{n'}({\bf x})h^{1/2}d^3x=\delta_{nn'}$. 
The principal quantum number ``$n$'' 
stands for  for quantum states of physical wave vectors $k_n$ 
and $k_0=0$ for the ground state \footnote{In Ref.~\cite{Parker1973}, the principal quantum number $n$ is the angular momentum number  ``$\ell=0,1,2,\cdot\cdot\cdot$'' and $Y_n({\bf x})=Y_{\ell,m}({\bf x})$ are the four-dimensional spherical harmonics for the closed Robertson-Walker metric. The ground state is $\ell =0$. 
}. 
The $A_n$ and $A_n^\dagger$ are time-independent annihilation and creation operators satisfying the commutation relation $[A_n^\dagger,A_n]=\delta_{n,n'}$. 
The time-separate equation for $\psi_n(t)$ is 
\begin{eqnarray}
\partial_t^2\psi_n(t) + \omega_n(t)^2\psi_n(t)=0,\quad 
\omega_n(t)^2=k^2_n+M^2,\label{timeeq}
\end{eqnarray} 
and Wronskian-type condition $\psi_n(t)\partial_t\psi^*_n(t) - \psi^*_n(t)\partial_t\psi_n(t)=i$. Expressing 
\begin{eqnarray}
\psi_n(t)\!&=&\!\frac{1}{(2V\omega_n)^{1/2}}\left(\alpha^*_n(t) e^{-i\int^t\omega_n dt}+\beta^*_n(t) e^{i\int^t\omega_n dt}\right)\label{alphabeta}
\end{eqnarray}
in terms of $\alpha_n(t)$ and $\beta_n(t)$, Equation (\ref{timeeq}) becomes
\begin{eqnarray}
\partial_t\alpha_n(t) &=& C_n e^{-2i\int^t\omega_n dt}\beta_n(t);\nonumber\\
 \partial_t\beta_n(t) &=& C_n e^{2i\int^t\omega_n dt}\alpha_n(t),
\label{eqalphabeta}
\end{eqnarray} 
and $|\alpha_n|^2-|\beta_n|^2=1$, where 
$C_n\equiv 3H\omega_n^{-2}[k_n^2/3 +M^2/2]$. 
In an adiabatic process for slowly time-varying $H=H_{\rm slow}$, namely quasi static case $H\approx {\rm const.}$, the particle state
$\alpha_n(0)=1$ and $\beta_n(0)=0$ evolve to $|\alpha_n(t)|\gtrsim 1$ and
$|\beta_n(t)|\not=0$. Positive and negative frequency modes get mixed, 
leading to particle productions of probability $|\beta_n(t)|^2\propto e^{-M/H_{\rm slow}}$. 

We will focus on studying particle production in non-adiabatic processes 
for rapidly time-varying $H_{\rm fast}$, $\alpha_n$ and $\beta_n$ in the 
ground state $n=0$ of the lowest lying massive mode $M\gg H$. 
First, we recall that Parker and Fulling introduced the 
transformation \cite{Parker1973}, 
\begin{eqnarray}
A_0=\gamma^*B + \delta B^\dagger,\quad B=\delta A^\dagger_0-\gamma A_0,
\label{bogo}
\end{eqnarray}
$[B,B^\dagger]=1$, and two mixing constants 
obeying $|\gamma|^2-|\delta|^2=1$. For a given $A_n$ 
and its Fock space, the state $|{\mathcal N}_{\rm pair}\rangle $ 
is defined by the conditions 
$A_{n\not=0}|{\mathcal N}_{\rm pair}\rangle=0$ and  
\begin{eqnarray}
B^\dagger B |{\mathcal N}_{\rm pair}\rangle ={\mathcal N}_{\rm pair}|{\mathcal N}_{\rm pair}\rangle.
\label{pairB}
\end{eqnarray}
The $B^\dagger$ and $B$ are time-independent creation and annihilation 
operators of the pair of mixed positive frequency $A_0$ particle and 
negative frequency $A_0^\dagger$ antiparticle. 
The state $|{\mathcal N}_{\rm pair} \rangle $ 
contains ${\mathcal N}_{\rm pair}=1,2,3,\cdot\cdot\cdot$ pairs, 
and $|{\mathcal N}_{\rm pair}=0 \rangle $ is the ground state 
of non-adiabatic interacting system of fast varying $H_{\rm fast}$ 
and massive pair production and annihilation \footnote{Discussions can be applied for fermion fields. Analogously, we discussed the back and forth processes of massive fermion 
and antifermion pairs production and annihilation in spacetime 
${\mathcal S}\Leftrightarrow \bar F + F$ in Refs.~\cite{Xue2019,Xue2020,Xue2020a}}.
It is a coherent superposition of states of particle and anti-particle pairs. In this coherent condensate state $|{\mathcal N}_{\rm pair} \rangle$ and ${\mathcal N}_{\rm pair}\gg 1$, neglecting higher mode 
$n\not=0$ contributions, they obtained the negative quantum pressure 
and positive quantum density of 
coherent pair field, see Eqs.~(59) and (60) of Ref.~\cite{Parker1973},
\begin{eqnarray}
p^{\rm fast}_{_M}&=&-\frac{M(2{\mathcal N}_{\rm pair }+1)}{2\pi^2 V}\Big\{{\rm Re}[\gamma^*\delta(|\alpha|^2+|\beta|^2)]\nonumber\\
&+&(2|\delta|^2+1){\rm Re}(\alpha^*\beta e^{2iMt})\Big\},\label{fastp}\\
\rho^{\rm fast}_{_M}&=&\frac{M(2{\mathcal N}_{\rm pair }+1)}{\pi^2 V}\Big\{{\rm Re}[\gamma\delta^*\alpha\beta)]\nonumber\\
&+&(|\delta|^2+1/2)(|\beta|^2+1/2)\Big\},
\label{fastrho}
\end{eqnarray}
where $\omega_{n=0}=M$, $\alpha_{n=0}=\alpha$ and $\beta_{n=0}=\beta$.
Pressure (\ref{fastp}) and density (\ref{fastrho}) were adopted for studying the avoidance of cosmic singularity in a curved Universe. 
In their sequent article \cite{Parker1974}, the authors confirm Eqs.~(\ref{fastp}) and 
(\ref{fastrho}) by studying the regularization of higher mode contributions to the energy-momentum tensor of a massive quantized field in closed, flat and hyperbolic spatial spaces. The natures of the massive coherent pair state $|{\mathcal N}_{\rm pair}\rangle$ (\ref{pairB}) of the pressure (\ref{fastp}) and density (\ref{fastrho}) are rather generic for non-adiabatic production of massive particles in curved spacetime. 

Following their approach for the ground state $k_n=0$, we arrive at the same quantum pressure (\ref{fastp}) and density (\ref{fastrho}). 
In our case, we consider the state (\ref{pairB}) 
as a coherent condensate state of very massive $M\gg H_{\rm slow}$ and 
large number ${\mathcal N}_{\rm pair}\gg 1$ pairs, 
and $M(2{\mathcal N}_{\rm pair}+1)$ in (\ref{fastp}) and (\ref{fastrho}) 
can be larger than the Planck mass
$m_{\rm pl}$. Therefore higher mode $(k_n\not=0)$ contributions could be neglected. Their regularization and corrections will be 
studied in future. In this article, we adopt (\ref{fastp}) and (\ref{fastrho}) as the fast components $\rho^{\rm fast}_{_M}$ and $p^{\rm fast}_{_M}$ in 
Eq.~(\ref{ffriedman}) to find their 
non-adiabatic back-reactions on fast components $H_{\rm fast}$ and 
$\rho^{\rm fast}_{_\Lambda}$.

Using negative $p^{\rm fast}_{_M}$ (\ref{fastp}) and positive definite 
$\rho^{\rm fast}_{_M}$ (\ref{fastrho}), we search for a solution of fast component equation (\ref{ffriedman}) and quantum fluctuating mode equations (\ref{eqalphabeta}) in the period $[-t,t]$ of the 
microscopic time $t\sim H^{-1}_{\rm fast}$. It is around the macroscopic time $t_{\rm slow}\sim H^{-1}_{\rm slow}$, when the slow components  $a_{\rm slow}$, $H_{\rm slow}$, $\rho^{\rm slow}_{_{M,\Lambda}}$ and $p^{\rm slow}_{_{M,\Lambda}}$ are valued, following 
the Friedman equations (\ref{sfriedman}).  The integrals 
$\int^t\omega_n dt$ are over the microscopic time $t$  characterised by 
the Compton time scale $1/M$. Its lower limit is $t=0$ by setting 
$t_{\rm slow}=0$ as a reference time, when $a_{\rm fast}(0)=0$,
\begin{eqnarray}
H_{\rm fast}(0)=\dot a_{\rm fast}/a_{\rm slow}=0;\quad \alpha(0)=1,\quad \beta(0)=0.
\label{initial}
\end{eqnarray}
The real value $\gamma^*\delta$ condition in Eqs.~(\ref{fastp}) 
and (\ref{fastrho}) leads to the time symmetry:
$a^{\rm fast}(t)=a^{\rm fast}(-t)$, $\alpha(t)=\alpha^*(-t)$ 
and $\beta(t)=\beta^*(-t)$ \cite{Parker1973}. When $t\leftrightarrow -t$, positive and negative frequency modes interchange. 
\comment{In Ref.~\cite{Parker1973}, $a_{\rm slow}=0$, $H_{\rm slow}=0$ (i.e.,
$H=H_{\rm fast}$, $a=a_{\rm fast}$) and a small spherical volume $V\sim (H_{\rm fast})^3$ at the cosmic origin were adopted for studying the avoidance of cosmic singularity for the $\rho_{_\Lambda}=0$ and curved Universe.} 
Here we use $a_{\rm slow}\not=0$, $H_{\rm slow}\not=0$ and co-moving radius $(Ha)^{-1}\approx (H_{\rm slow}a_{\rm slow})^{-1}$ of Hubble volume $V\sim H_{\rm slow}^{-3}$.

\begin{figure*}[t]
\centering
\begin{center}
\includegraphics[height=5.5cm,width=7.8cm]{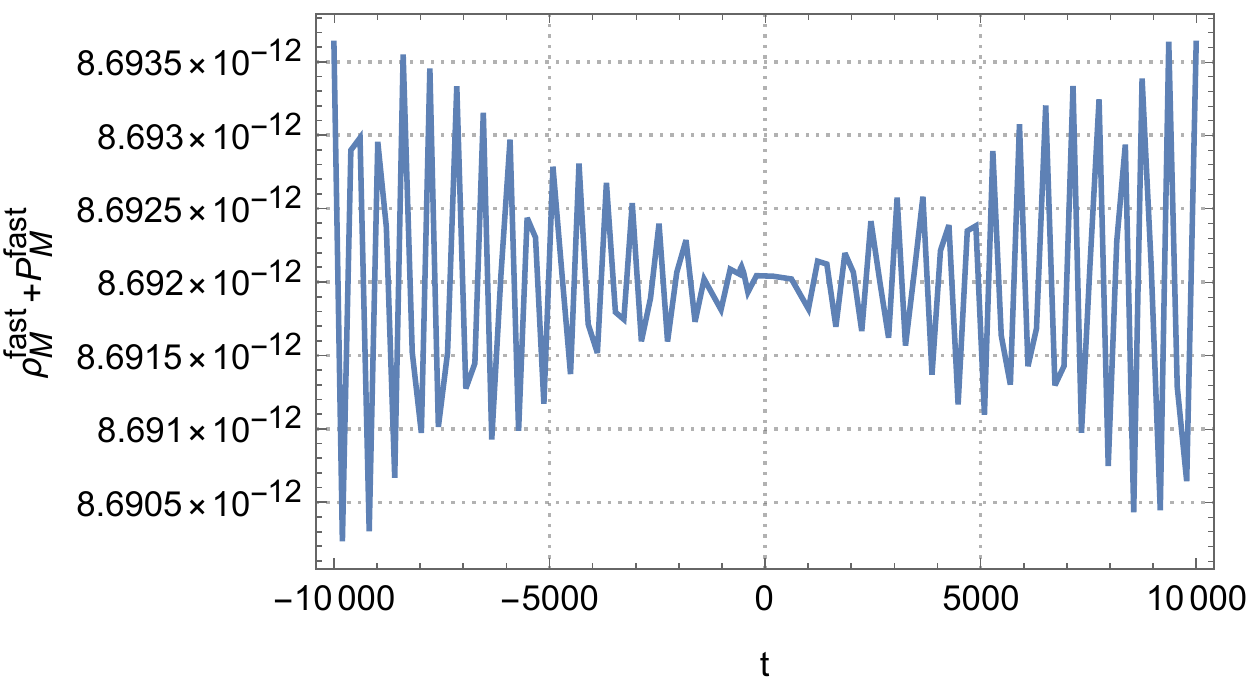}
\caption{The quantum pair density and pressure oscillations in time are shown using  $M=10^{-2} m_{\rm pl}$, $H_{\rm slow}= 10^{-5} m_{\rm pl}$, 
${\mathcal N}_{\rm pair}=10^{8}$ and $\delta= 1$. The Planck unit $m_{\rm pl}=(1/8\pi G)^{1/2}=1$ is adopted for presenting numerical results, namely the time ``$t$'' is in the unit of the Planck time, the quantum pair density 
$\rho^{\rm fast}_{_M}$ and pressure $p^{\rm fast}_{_M}$ are in the unit of the Planck density. It shows that $\rho^{\rm fast}_{_M}$ and $p^{\rm fast}_{_M}$ are not small, but their oscillating amplitudes $\delta\rho^{\rm fast}_{_M}/\rho^{\rm fast}_{_M}$ and $\delta p^{\rm fast}_{_M}/p^{\rm fast}_{_M}$ are about ${\mathcal O}(10^{-3})$. For a long time $t> 10^4$, the coherent oscillations approach stable configurations in time, and amplitude damping effects appear. For more details and figures, see Fig.~\ref{detailosci1} 
in Appendix of Supplemental Material. 
}\label{osci}
\end{center}
\vspace{-2em}
\end{figure*}

In microscopic time $t$ of unit $m_{\rm pl}^{-1}$, we numerically solve ``microscopic'' Equations (\ref{ffriedman}), (\ref{eqalphabeta}),  
(\ref{fastp}) and (\ref{fastrho}) 
that are non-linearly coupled equations at time scale ${\mathcal O}(M^{-1})$. In addition to massive pairs production and coherent state (\ref{pairB}) formation, we find (Fig.~\ref{osci}) that the system undergoes {\it quantum pair oscillation}, 
namely the quantum pressure $p^{\rm fast}_{_M}$ (\ref{fastp}) and 
density $\rho^{\rm fast}_{_M}$ (\ref{fastrho}) coherently oscillating with 
$H_{\rm fast}$ and $\rho^{\rm fast}_{_\Lambda}$. Figure \ref{osci} shows results for 
$C_0 = (3/2)H_{\rm fast}$ and verified condition $|\alpha|^2-|\beta|^2=1$. 
In the quantum period of microscopic time $t$, the negative quantum pressure $p^{\rm fast}_{_M}< 0$ and ``microscopic'' ${\mathcal O}(M^{-1})$ 
back-reaction effects lead to the 
quantum pair oscillation characterised by the frequency $\omega=M$ of massive quantised pair fields. 
The positive quantum pair density $\rho^{\rm fast}_{_M}> 0$  
indicates particle creations without $e^{-M/H}$ suppression. 
It is consistent with increasing Bogoliubov 
coefficient $|\beta(t)|^2$ that mixes positive and negative energy modes. 
The sum $\rho^{\rm fast}_{_M}+ p^{\rm fast}_{_M} >0$ is positive definite, leading to the decreasing $H_{\rm fast}(t)$ (\ref{ffriedman}). As a consequence,   
for time $t>0$, the fast components $H_{\rm fast}$ 
and $\rho^{\rm fast}_{_\Lambda}$ decrease in time, in order for pair production. Whereas for time $t<0$, $H_{\rm fast}$ and $\rho^{\rm fast}_{_\Lambda}$ increases, due to pair annihilation. The small $a_{\rm fast}(t)$ varies around $a_{\rm slow}$ at $t_{\rm slow}\equiv 0$. Note that $p^{\rm fast}_{_M}$ (\ref{fastp}) and $\rho^{\rm fast}_{_M}$ (\ref{fastrho})
represent the quantum pressure and density of massive coherent pair 
state (\ref{pairB}) in short quantum time sales ${\mathcal O}(1/M)$. They do not follow the usual 
equation of the state of classical matter.

We numerically solve non-linearly back-reacting equations (\ref{ffriedman}),
(\ref{eqalphabeta}), (\ref{fastp}) and (\ref{fastrho}) for fast components at the microscopic scale. As a result, one of our findings in this article is the high-frequency ${\mathcal O}(M^{-1})$ oscillation of quantum pair state's pressure 
$p^{\rm fast}_{_M}$ and density $\rho^{\rm fast}_{_M}$, coherently with the oscillations of the fast components $H_{\rm fast}$, 
$\dot H_{\rm fast}$ and $\rho_{_\Lambda}^{\rm fast}$, 
see Fig.~\ref{osci}, Figs.~\ref{detailosci1} and \ref{detailosci0} in supplemental material. By showing the highly non-adiabatic nature of pair-production processes, 
we present the novel dynamics and quasi-classical state of collective 
oscillations, in addition to the coherent state $|{\mathcal N}_{\rm pair}\rangle$ 
(\ref{pairB}) of massive pair production \cite{Parker1973}. Such quantum pair oscillation phenomenon is dynamically analogous 
to the quasi-classical plasma oscillation of electron-positron pair production in 
an external electric filed $E$ \cite{Kluger1991} and pair production rate 
is not exponentially suppressed by $e^{-\pi M^2/E}$ \cite{Ruffini2010}. 
This analogy motivates us to model the quantum coherent pair state  
$|{\mathcal N}_{\rm pair}\rangle $ (\ref{pairB}) and oscillating dynamics (Fig.~\ref{osci}) as a quasi-classical plasma state
of effective energy density and pressure to study their impacts on the Friedman equation (\ref{sfriedman}) in the $\tilde\Lambda$CDM scenario. 

\section{Massive pair plasma state}\label{mpp}

As shown in Fig.~\ref{osci}, massive pair quantum pressure $p^{\rm fast}_{_M}$ (\ref{fastp}) and density $\rho^{\rm fast}_{_M}$ (\ref{fastp})  can be significantly large and rapidly oscillate with the fast components $H_{\rm fast}$ and $\rho^{\rm fast}_{_\Lambda}$ (\ref{ffriedman}) 
in microscopic time and space. Their oscillating amplitudes are not dampen in time, and it is therefore expected to form 
{\it a massive pair plasma state} in a long macroscopic time. 
However, to study their effective impacts on the classical Friedman equations (\ref{sfriedman}) evolving in macroscopic time and space, we have to discuss two problems coming from scale difference 
$M\gg H$. First, it is impossible to even numerically integrate slow and fast component coupled equations (\ref{sfriedman},\ref{ffriedman}) due to their vastly different time scales. On this aspect, we consider their 
non-vanishing averages $\langle\cdot\cdot\cdot\rangle$ over the microscopic period in time. Figure \ref{osci} shows $\langle\rho^{\rm fast}_{_M}+ p^{\rm fast}_{_M}\rangle$ and other averages of fast oscillating components do not vanish. Second the spatial dependence of pair quantum pressure $p_{\rm fast}$ (\ref{fastp}) and density $\rho_{\rm fast}$ (\ref{fastp}) are unknown, since they are obtained 
by using the vacuum expectation value of field $\Phi({\bf x},t)$ energy-momentum tensor over entire space. For the case $M\gg H_{\rm slow}$, 
the Compton length $M^{-1}$ of ground state $n=0$ is much smaller than the Hubble horizon $H^{-1}_{\rm slow}$. Therefore, the massive coherent pair state (\ref{pairB}-\ref{fastrho}) and quantum plasma oscillation of Fig.~\ref{osci} well localise inside the Hubble sphere. We speculate that their location should be near to the Horizon rather than at the centre, because of isotropic homogeneity extending up to the horizon.  

Based on these considerations and non-vanishing averages 
of fast oscillating components (Fig.~\ref{osci}) over microscopic 
time period, we assume the formation of massive pair plasma state at macroscopic time scale ${\mathcal O}(H^{-1}_{\rm slow})$. We describe such macroscopic state as a perfect fluid state of effective number $n^H_{_M}$ and 
energy $\rho^H_{_M}$ densities as,
\begin{eqnarray}
\rho^H_{_M} \equiv  2\chi  m^2 H^2_{\rm slow},\quad n^H_{_M} \equiv   \chi  m H^2_{\rm slow};\quad m^2 \equiv 
\sum_fg_d^fM^2_f,
\label{apdenm}
\end{eqnarray}
and pressure $p^H_{_M}=\omega^H_{_M}
\rho^H_{_M}$. The $\omega^H_{_M}\approx 0$ for $m\gg H_{\rm slow}$
and its upper limit is $1/3$. The introduced mass parameter $m$ represents possible particle masses $M_f$, degeneracies $g_d^f$ 
and the mixing coefficient $\delta$ (\ref{bogo}). 
The degeneracies $g_d^f$ plays the same role of pair number ${\mathcal N}_{\rm pair}$ in 
Eq.~(\ref{fastrho}), namely $\sum_f g_d^f\approx (2{\mathcal N}_{\rm pair}+1)$. We explain 
the reasons why the densities (\ref{apdenm}) 
are proportional to $\chi m H^2_{\rm slow}$, rather than $H^3_{\rm slow}$ 
from the entire Hubble volume $V$. The ``surface area'' factor $H^2_{\rm slow}$ 
is attributed to the 
spherical symmetry of Hubble volume. The ``radial size'' factor $\chi m$ comes from 
the layer width $\lambda_m $ introduced as an effective parameter to
describe the properties (i) for $m\gg H_{\rm slow}$ the massive pair plasma
is localised as a spherical layer and (ii) its radial width $\lambda_m < H^{-1}_{\rm slow}$
depends on the massive pair plasma oscillation dynamics \footnote{It may also include self-gravitating dynamics, due to pair plasma are very massive.}, 
rather than the $H_{\rm slow}$ dynamics govern by the Friedman equations (\ref{sfriedman}). 
The width parameter $\chi$ expresses the layer width $\lambda_m = (\chi m)^{-1} \gg 1/m$ 
in terms of the effective Compton length $1/m$,
\begin{eqnarray}
\lambda_m=(\chi m)^{-1} < H^{-1}_{\rm slow},\quad  1\gg \chi > (H_{\rm slow}/m).
\label{chi}
\end{eqnarray} 
Because parameters $m$ and $\chi m$ represent time-averaged values over 
fast time oscillations of massive pair plasma state, we consider $m$ and 
$\chi m$ as approximate constants in slowly varying macroscopic 
time. However, the typical $m$ and $\chi m$ values should be 
different for Universe evolution epochs, since the fast-component 
equations for massive pair productions and oscillations depend on 
the $H_{\rm slow}$ value, see Sec.~\ref{qppo}. 
Their values have to be fixed by observations. On the other hand, in the approximation without separating fast and slow components, we have consistently obtained the mean density $n^H_{_M} \approx \chi  m H^2$ (\ref{apdenm}) and $\chi\approx 1.85\times 10^{-3}$ by studying massive fermion pair productions in an exact De Sitter spacetime of $H={\rm const.}$~and scaling factor $a(t)=e^{iHt}$ \cite{Xue2019, Xue2020}. 

We have to point out the following. (i) The pressure $p^H_{_M}$ and density $\rho^H_{_M}$ (\ref{apdenm}) are effective descriptions of the massive pair plasma state in macroscopic scales. It results from the coherence condensation state (\ref{pairB},\ref{fastp},\ref{fastrho}) and oscillating dynamics (Fig.~\ref{osci}) in microscopic scales. (ii) They play the role of ``slow'' components contributing to the ``macroscopic'' ${\mathcal O}(H^{-1}_{\rm slow})$ Friedman equations (\ref{friedman}) or (\ref{sfriedman}). It means that in the Friedman equations (\ref{sfriedman}), the matter density $\rho^{\rm slow}_{_M}$ and pressure $p^{\rm slow}_{_M}$ contains contributions from (a) the normal matter state of pressure and density 
and (b) the massive pair plasma state of pressure and density $p^H_{_M}=\omega^H_{_M}\rho^H_{_M}$. These two sets may interact with each other. We shall study the massive pair plasma state effects on each epoch of the Universe's evolution. Here we start to study its effects on inflation. Henceforth sub- and super-scripts ``slow'' are dropped.

\section{Massive pair plasma state effect on inflation}\label{mfeq}

To start this section, we recall the Ref.~\cite{Starobinsky1978}, 
showing that the massive pair state $|{\mathcal N}_{\rm pair}\rangle$ (\ref{pairB})of the large occupation number ${\mathcal N}_{\rm pair}\gg 1$ is a quasi-classical state equivalently to the FLRW model filled by a massive classical scalar field. The author obtained the analytical solution of the slowly evolving quasi-de Sitter stage for inflation. Here, we study the state $|{\mathcal N}_{\rm pair}\rangle$ in the $\tilde\Lambda$CDM scenario by showing the fast oscillating components $H_{\rm fast}$ and $\rho_{_\Lambda}^{\rm fast}$ produce the massive particle pairs. Moreover, the massive pair state's energy density and pressure in coherent oscillation with $H_{\rm fast}$ and $\rho_{_\Lambda}^{\rm fast}$ can thus form a quasi-classical and massive plasma
state (\ref{apdenm}). We will study at the macroscopic time scale 
${\mathcal O}(H^{-1})$ how the time-varying cosmological term 
$\rho_{_\Lambda}^{\rm slow}$ derives the inflation and how the quasi-classical and massive plasma state (\ref{apdenm}) slows down the inflation. 
It is different from the inflation model of a massive scalar field of the potential $V(\phi)\propto M^2\phi^2$ in the FLRW metric.

In this section, we show that the inflation is driven by the cosmological term $\rho_{_\Lambda}(t)$ (gravitationally repulsive) and it is slowed down by the massive pair plasma state (\ref{apdenm}) (gravitationally attractive). 
The latter is formed at the expense of the former energy.
Suppose that during inflation the normal matter state 
of pressure and density 
is absent, and 
only massive pair plasma state of pressure and density 
$p^H_{_M}=\omega^H_{_M}\rho^H_{_M}$  (\ref{apdenm}) is present. 
The ``macroscopic'' ${\mathcal O}(H^{-1})$ Friedman equations (\ref{friedman}) become 
\begin{eqnarray}
H^2&=&\frac{8\pi G}{3}(\rho_{_\Lambda}+\rho^H_{_M}),\nonumber\\
\dot H&=&-\frac{8\pi G}{2}(\rho^H_{_M}+p^H_{_M}).
\label{xfriedman}
\end{eqnarray}
These Equations (\ref{xfriedman}), time-varying ``dark energy'' 
$\rho_{_\Lambda}=\tilde\Lambda/(8\pi G)$, massive plasma state
$\rho^H_{_M}$ and $p^H_{_M}$ (\ref{apdenm}) give a ``macroscopic''  back-reacting system  at the scale ${\mathcal O}(H^{-1})$, yielding a slowly time-decreasing $H$ for the quasi-de Sitter phase (\ref{apps0}) discussed below. 
This should be differentiated from the ``microscopic'' ${\mathcal O}(M^{-1})$ back-reacting system of Eqs.~(\ref{ffriedman}), (\ref{eqalphabeta}), (\ref{fastp}) and (\ref{fastrho}), yielding the quantum pair coherent oscillation 
discussed before. It is a difficult task to analyse ${\mathcal O}(M^{-1})$ and ${\mathcal O}(H^{-1})$ dynamics numerically since two scales $M\gg H$ are very different. It is the reason why we split the fast ${\mathcal O}(M^{-1})$  components from the slow ${\mathcal O}(H^{-1})$ components, and introduce at the scale ${\mathcal O}(H^{-1})$ the massive pair plasma state of effective density $\rho^H_{_M}$ and pressure $p^H_{_M}$ (\ref{apdenm}). They are microscopic time averages over fast components (\ref{fastp},\ref{fastrho}) and contribute to slow components in Friedman equation (\ref{xfriedman}).

In the inflation epoch, the time-varying cosmological term $\rho_{_\Lambda}$ is dominant over the massive pair plasma state $\rho^H_{_M}$, e.g., $\rho_{_\Lambda}\gg \rho^H_{_M}$. The former derives the inflation, while the latter slowly slows it down. 
Assuming initial values of ``cosmological constant'' $\tilde\Lambda(0)=\Lambda$ and 
$H(0)= (\Lambda/3)^{1/2}$ \footnote{We expect the initial cosmological constant $\Lambda$ value to be in the range between the GUT scale ($\sim 10^{15}$ GeV) 
and the Planck scale. It is not an issue here to discuss the quantum-gravity origin of cosmological constant $\Lambda$, which possibly represents the correlation length $\xi$ (characteristic scale) of quantum gravity field theory, 
$\Lambda\sim \xi^{-2}$ \cite{Xue2010,Xue2012,Xue2009,Xue2015}, analogously to the scale $\Lambda_{\rm QCD}$ of the quantum chromodynamics field theory.}, Eqs.~(\ref{xfriedman}) show that the scalar factor $a\sim \exp (\Lambda/3)^{1/2}t$ is exponentially inflated in time if the massive pair plasma state is absent $\rho^H_{_M}=p^H_{_M}\equiv 0$. 
As the consequence of the nontrivial massive pair plasma state 
($\rho^H_{_M}\not=0, p^H_{_M}\approx 0$) and its back reaction on $H$ via $\dot H <0$ of Eq.~(\ref{xfriedman}), $H$ and $\tilde\Lambda$ decrease in time, become dynamically time dependent. Thus inflation is slowed down to its end. 


\section{Inflation and tensor-to-scalar ratio}

As the macroscopic time $t$ varies at the scale $H^{-1}$, what is the rate of pair production in connection with the massive pair plasma state density (\ref{apdenm}) changing and contributing to the matter density. To quantitatively describe these dynamics, 
we estimate the total number of particles produced inside the Hubble sphere $N\approx n^H_{_M}H^{-3}/2$ and mean pair production rate w.r.t.~macroscopic time variation $dt$,
\begin{eqnarray}
\Gamma_M \approx \frac{dN}{2\pi dt}\approx \frac{\chi m}{4\pi} \epsilon,\quad \epsilon\equiv -\frac{\dot H}{H^2}.
\label{prate}
\end{eqnarray}
Here we neglect 
the back-reactions of slow time-varying components $H$, $\rho_{_{\Lambda,M}}$ and 
$p_{_{\Lambda,M}}$ on fast components $H_{\rm fast}$, $\rho^{\rm fast}_{_M}$ and $p^{\rm fast}_{_M}$. The modified Friedman equations (\ref{xfriedman}) and rate (\ref{prate}) are basic equations to quantitatively describe inflation, effective mass $m$ and width 
$\chi$ parameters are fixed by observations.

In inflation, the $H$ is larger than the mean pair production rate 
$H>\Gamma_M$, Equations (\ref{xfriedman}) are governed by cosmological term $\rho_{_\Lambda}$ and induced massive pair plasma state 
of density $\rho^H_{_M}$ (\ref{apdenm}) and pressure $p^H_{_M}\approx 0$. 
From Eqs.~(\ref{apdenm}) and (\ref{xfriedman}), we analytically obtain the inflationary solution of slowly decreasing $H$ (slow-rolling dynamics)
\begin{eqnarray}
H \approx H_*(a/a_*)^{-\epsilon},\quad \epsilon\approx \chi (m/m_{\rm pl})^2\ll 1,
\label{apps0}
\end{eqnarray} 
where $a_*$ and $H_*$ are the characteristic inflation scale   
corresponding to the interested quantum modes 
of pivot scale $k_*$ 
crossed the horizon $(c_sk_*=H_*a_*)$ for CMB observations. 
Here, the interested quantum modes refer to the primordial curvature perturbations of the standard scenario. We will discuss separately the possibly interesting quantum modes of quantum pair coherent oscillations presented in the previous section \ref{qppo}. Therefore,
the scalar, tensor power spectra and their ratio read \cite{Baumann2015}
\begin{eqnarray}
\hspace{-1.5em}\Delta^2_{_{\mathcal R}} 
\!&=&\! \frac{1}{8\pi^2}\frac{H^2_*}{m^2_{\rm pl}\,\epsilon\,c_s},
~\Delta^2_h 
\!=\! \frac{2}{\pi^2}\frac{H^2_*}{m^2_{\rm pl}};~
r\!\equiv\! \frac{\Delta^2_h}{\Delta^2_{_{\mathcal R}}}=16\,\epsilon\, c_s,
\label{ps}
\end{eqnarray} 
where the time-dependent background sound velocity $c_s< 1$, 
and the spectra index $n_s\approx 1-2\epsilon$ at the leading order of scale-invariance deviations. Based on two CMB observational values 
at the pivot scale $k_*=0.\,05\, ({\rm Mpc})^{-1}$ \cite{Aghanim2020}: 
(i) the spectral index
$n_s\approx 0.965$, 
from Eq.~(\ref{apps0}) we obtain
\begin{equation}
\epsilon\approx \chi (m_*/m_{\rm pl})^2\lesssim (1-n_s)/2\approx 0.0175,
\label{m*}
\end{equation}
and the $m_*$ is the mass scale (\ref{apdenm}) corresponding to 
the pivot scale; 
(ii) the scalar amplitude $A_s=\Delta^2_{_{\mathcal R}}(k_*)
\approx  2.1\times 10^{-9}$, 
Equation (\ref{ps}) gives
\begin{equation}
H_*=3.15\times 10^{-5}\,(r/0.1)^{1/2}m_{\rm pl}. 
\label{H*}
\end{equation}
As a result, the energy-density ratio of pair plasma and cosmological term densities is
\begin{eqnarray}
\frac{\rho^H_{_M}}{\rho_{_\Lambda}}\Big|_{H_*}
\approx \frac{2\chi (m_*H_*)^2}{3(m_{\rm pl}H_*)^2}
=\frac{2}{3}\chi \left(\frac{m_*}{m_{\rm pl}}\right)^2\approx 1.17\times 10^{-2},
\label{ratio*}
\end{eqnarray}
and $H^2_*\approx \rho^*_{_\Lambda}/(3m^2_{\rm pl})$. 

\begin{figure}
\centering   
\begin{center}
\includegraphics[height=6.5cm,width=7.8cm]{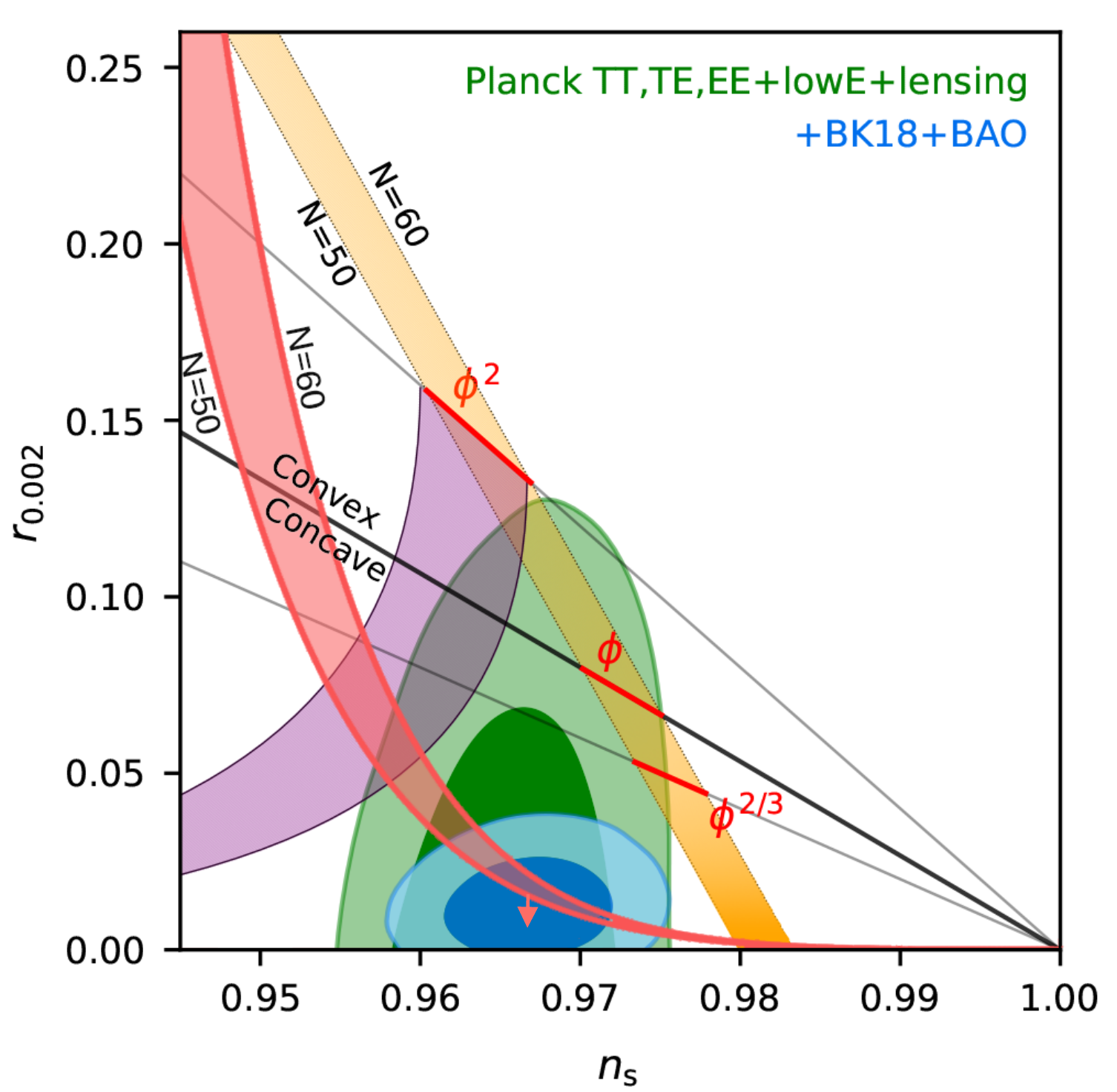}
\vspace{-1em}
\caption{On the Figure 5 of Ref.~\cite{Ade2021},  
the upper limit (\ref{endr}) is plotted for $\chi=10^{-3}$.  
The red zone bound by the $e$-folding number $N_{\rm end}=N=50,60$ curves 
agrees with the blue constraint zone. The constraint on the tensor-to-scalar ratio $r$ also agrees with the upper limit $r<0.044$ from the recent observation \cite{Tristram2021}. 
The Figure 5 of Ref.~\cite{Ade2021} comes from Figure 28 of Ref.~\cite{Akrami2020} and Figure 8 of Ref.~\cite{Aghanim2020}. Their figure captions indicate the inflation models studied. For example, the yellow region shows the loci
of approximately constant $e$-folding number $N$, assuming simple
$V(\phi)\propto (\phi/m_{\rm pl})^p$ single-field inflation. It shows that the red zone of the $\tilde\Lambda$CDM is distinct from the constrained zones of other inflation models.
}
\label{ns-rplot}
\end{center}
\vspace{-2em}
\end{figure}

The inflation slows down and 
eventually ends at $a=a_{\rm end}$ and $H=H_{\rm end}$, 
\begin{eqnarray}
H_{\rm end}=H_*\exp -(\epsilon\, N_{\rm end}),
\label{hend}
\end{eqnarray}
where  
$N_{\rm end}=\ln \left(a_{\rm end}/a_*\right)$ 
is the $e$-folding numbers from the inflation scale $H_*$ 
to the inflation ending scale 
$H_{\rm end}$. It can be preliminarily determined by the inflationary 
rate being smaller than the mean pair-production rate  
namely
\begin{eqnarray}
H_{\rm end}< \Gamma_M =(\chi m_*/4\pi)\,\epsilon.
\label{infend}
\end{eqnarray}
However, 
this inequality provides the upper bound on $H_{\rm end}$, whose value 
should be calculated by studying the dynamical transition from
inflation to reheating.
Using Eqs.~(\ref{hend}) and (\ref{infend}), we give the upper limit on the 
tensor-to-scalar ratio 
$r$ in terms of the $e$-folding numbers $N_{\rm end}$,
\begin{eqnarray}
r< 1.01&\times& 10^8 \left(\frac{\Gamma_M}{m_{\rm pl}}\right)^2 e^{2\epsilon N_{\rm end}}\nonumber\\
&=& 7.97\times 10^4 \chi  (1-n_s)^3e^{(1-n_s) N_{\rm end}}
\label{endr}
\end{eqnarray}
where
$\epsilon =\chi (m_*/m_{\rm pl})^2=(1-n_s)/2$ (\ref{m*}) is used. 
Non-vanishing $\chi$ implies $r\not =0$. In Fig.~\ref{ns-rplot}, we plot 
the upper limit (\ref{endr}) compared with data and other inflation models. The range of width parameter $\chi$ values 
is discussed in Eq.~(\ref{chi}). The inequalities $\lambda_m\gg 1/m$ 
and $m\gg H$ implies $\chi\ll 1$.  
For Fig.~\ref{ns-rplot} and calculations below, we chose the reference value 
$\chi\sim 10^{-3}$  at the same order of $\chi\approx 1.85\times 10^{-3}$ that we approximately obtained for massive fermion pair productions in an exact De Sitter spacetime \cite{Xue2019, Xue2020}.
\comment{The observationally 
fixed values $\chi m^2\sim 10^{-2}$ (\ref{m*}) and $\chi\sim 10^{-3}$ imply 
$\chi {\mathcal N}_{\rm pair}\sim 10^{-2}(m_{\rm pl}/M)^2$, indicating 
the produced pair number ${\mathcal N}_{\rm pair}\sim 10^5$ within a layer width 
$(\chi M)^{-1}$, if pair mass $M/m_{\rm pl}\sim 10^{-2}$ and $M/H_{\rm slow}\sim 10^3$. It is the parameter set for Figures \ref{???}.
} 

From Eq.~(\ref{hend}), the inflation ending scale $H_{\rm end}$ is given by
\begin{eqnarray}
H_{\rm end}
&\approx &H_*e^{-(1-n_s)N_{\rm end}/2} \approx(0.42,0.35)H_*, 
\label{Hend}
\end{eqnarray}
for $N_{\rm end}= (50,60)$ and $r=(0.02,0.028)$.  
It shows small $H$-variation 
\begin{eqnarray}
H^2_{\rm end} = \frac{ \rho^{\rm end}_{_\Lambda}  + \rho^{H\rm end}_{_M}}{3m^2_{\rm pl}}\gtrsim \frac{ \rho^{\rm end}_{_\Lambda}}{3m^2_{\rm pl}}; \quad 
\frac{\rho^{H\rm end}_{_M}}{\rho^{\rm end}_{_\Lambda}}\ll 1,
\label{ratioend}
\end{eqnarray}
and $\rho^{\rm end}_{_\Lambda}\approx 3
m^2_{\rm pl}H^2_{\rm end}$. Equations (\ref{ratio*}) and (\ref{ratioend}) imply the time-varying $\Lambda(t)\propto H^2$ ``area law'' in inflation. 

We would like to point out that the quasi-de Sitter phase (slow-rolling dynamics) for inflation undergoes when $\rho_{_\Lambda}$ and $H$ slowly decrease in time. In this epoch, $H>\Gamma_M$, the massive plasma state energy density $\rho^{H}_{_M}$ is much smaller than $\rho_{_\Lambda}$ (\ref{ratio*}), 
and slowly increases in time. Therefore $\rho^{H}_{_M}$ 
back-reaction on $\rho_{_\Lambda}$ is small, leading to 
slowly time-decreasing $\rho_{_\Lambda}$ that 
predominately governs the $H$ evolution, slowly decreasing in time from $H_*$ and $H_{\rm end}$ (\ref{Hend}). At the inflation end $H \lesssim \Gamma_M$ and the transition to $H < \Gamma_M$, the quantum pair production and oscillation play an important role. The ``dark-energy'' density $\rho_{_\Lambda}$ decreases rapidly and converts to the energy density $\rho^{H}_{_M}$ of massive pairs. As a result, 
$\rho^{H}_{_M}$ becomes comparable with, then predominates over $\rho_{_\Lambda}$, e.g., a matter-dominate episode 
$\rho^{H}_{_M}\gg \rho_{_\Lambda}$ and 
$\rho_{_\Lambda}\rightarrow 0$ ($\tilde\Lambda \rightarrow 0$). 
Moreover, massive pairs decay to light particles and decay rate $\Gamma^{^{\rm de}}_M > H$. 
The massive pairs' energy density converts to the radiation energy density $\rho_{_R}$, leading to the radiation-dominant reheating. 
The situations are similar to discussions 
in Refs.~\cite{Kofman1994, Kofman1997}. On this issue, 
we present in Ref.~\cite{Xue2020a} preliminary analysis and will publish
lengthy calculations and final results in a separate article.

\section{Comparison with other inflation models}

We compare and contrast the $\tilde\Lambda$CDM scenario to inflation models with scalar field potentials. 
We emphasise that the quantum scalar field $\Phi$ and equations (\ref{qfield}-\ref{fastrho}) describe the state and dynamics of massive pair productions in the Friedman Universe with dark energy density $\rho_{_\Lambda}$. They are not inflation field $\phi$, potential $V(\phi)$,
$3m_{\rm pl}^2H^2=\dot\phi^2/2 +V(\phi)$, 
$\rho_\phi=\dot\phi^2/2 + V(\phi)$ and $p_\phi=\dot\phi^2/2 - V(\phi)$ in inflation models. However, we can find some correspondences between inflation models and the present scenario (\ref{apdenm},\ref{xfriedman}) 
\begin{eqnarray}
\dot\phi^2 \Leftrightarrow \rho^{H}_{_M} 
+ p^{H}_{_M}\approx \rho^{H}_{_M}, \quad 
V(\phi) \Leftrightarrow  \rho_{_\Lambda} + (\rho^{H}_{_M} - p^{H}_{_M})/2
\approx \rho_{_\Lambda}.
\label{corr}
\end{eqnarray}
The slow-roll condition  $V(\phi)\gg \dot\phi^2/2$ corresponds 
to $\rho_{_\Lambda} \gg \rho^{H}_{_M}$ for 
$\rho^H_{_M}\approx (2\chi m^2/3m^2_{\rm pl}) \rho_{_\Lambda}$. It leads to $\dot \rho_{_\Lambda} \Leftrightarrow \dot V=\dot\phi V^\prime$ and $\dot \rho^H_{_M} \Leftrightarrow (1/2)d(\dot\phi^2)/dt =\dot\phi\ddot\phi $. As a result, the second equation in (\ref{xfriedman}) corresponds to the classical equation of motion for $\phi$: $\ddot\phi+3H\dot\phi +V^\prime(\phi)=0$. These correspondences imply that the $\Lambda$CDM scenario (\ref{apdenm},\ref{xfriedman}) could be effectively expressed in terms of inflation field $\phi$ and peculiar potentials $V(\phi)$. 

We have to mention the pioneer $R+R^2$ inflation model \cite{Starobinsky1980}, which agrees with the observational constraints on the spectral index $n_s$ and tensor-to-scalar ratio $r$. 
It is worthwhile to see the connection 
between the $\tilde\Lambda$CDM scenario 
and the $R+R^2$ model from the viewpoints of the asymptotic safety \cite{Weinberg2010} and cosmological observations \cite{Starobinsky1998}. The non-local UV-complete gravitational and particle field theory of higher derivatives \cite{Koshelev2016} or holonomy fields along a loop [Eq.~(133)] of Ref.~\cite{Xue2010} can have fixed points \cite{Weinberg2010, Xue2012, Xue2015}. Their
scaling domains can realise the effective and quasi-classical field theory of gravity and particles. One has to investigate, in agreement with observations, the 
following issues. If there is one scaling domain for the inflation dynamics. What are effectively relevant operators, $\langle T^{\mu\nu}_M\rangle$, $\tilde\Lambda$, $R$ and $R^2$. What are scaling laws for these operators as the cosmological scale changes? How we use an effective potential approach to describe the dynamics of these relevant operators. These are subjects for future studies.

\section{Singularity-free pre-inflation and large-scale anomaly}
It is worthwhile to mention the results for pre-inflation in the $\tilde\Lambda$CDM scenario. In the pre-inflation, when the Hubble scale 
$H\sim H_{\rm fast}=\dot a_{\rm fast}/a_{\rm fast}$ and all slow components are zero, namely $a_{\rm slow}=0, H_{\rm slow}=0$, $p_{_{M,\Lambda}}^{\rm slow}=0$ and $\rho_{_{M,\Lambda}}^{\rm slow}=0$, 
see Sec.~\ref{int}. The $H_{\rm fast}$ and $a_{\rm fast}$ dynamical evolution are govern by the fast components $\rho_{_\Lambda}^{\rm fast}$, $\rho_{_M}^{\rm fast}$ and $p_{_M}^{\rm fast}$. 
The Friedman equations (\ref{friedman}) become
\begin{eqnarray}
H^2_{\rm fast}=\frac{8\pi G}{3}(\rho^{\rm fast}_{_\Lambda}+\rho_{_M}^{\rm fast}),~\dot H_{\rm fast} =-\frac{8\pi G}{2}(\rho_{_M}^{\rm fast} + p_{_M}^{\rm fast}),\label{pfriedman}
\end{eqnarray}
with $\rho_{_M}^{\rm fast}$ (\ref{fastp}) and $p_{_M}^{\rm fast}$ (\ref{fastrho})
in a spherical Hubble volume $V\sim H^{-3}_{\rm fast}$. The initial values are (\ref{initial}), but $H^2_{\rm fast}(0)\approx (m^{-2}_{\rm pl}/3)\rho^{\rm fast}_{_\Lambda}(0)\not=0$ and $a_{\rm fast}(0)\not=0$, due to nontrivial cosmological term $\rho^{\rm fast}_{_\Lambda}(0)=\Lambda/(8 \pi G)$ and $\Lambda\sim m_{\rm pl}^2/a_{\rm fast}^2(0)$. The $\Lambda$ value is about the Planck scale, attributed to the nature of quantum gravity. 
Numerically integrating Eqs.~(\ref{eqalphabeta}), (\ref{fastp}), (\ref{fastrho}) and (\ref{pfriedman}), we show that quantum pair production and oscillation do not decrease 
the scale factor $a_{\rm fast}(t)$, which instead exponentially increases, 
leading to inflation. It concludes that the Universe does not
contract to a spacetime singularity of infinite density and gravity. The results
show that the weak energy condition of 
$\rho^{\rm fast} = \rho^{\rm fast}_{_M} + \rho^{\rm fast}_{_\Lambda}>0$ and $\rho^{\rm fast} + p^{\rm fast} =\rho^{\rm fast}_{_M} 
+ p^{\rm fast}_{_M}>0$ is satisfied, 
but the strong energy condition $\rho^{\rm fast} + 3p^{\rm fast} =
\rho^{\rm fast}_{_M} + 3p^{\rm fast}_{_M} -2 \rho^{\rm fast}_{_\Lambda}  
>0$ is violated for details see Fig.~\ref{detailosci0} 
in Supplemental Material.

Using Friedman Equations (\ref{xfriedman}) and $\epsilon$-rate 
$\epsilon =-\dot H/H^2$ (\ref{prate}), we recast the scalar spectrum (\ref{ps}) of primordial curvature perturbations as 
\begin{eqnarray}
\Delta^2_{_{\mathcal R}} (k) 
&\approx & \frac{1}{12\pi^2}\frac{\rho_{_\Lambda}}{c_s\chi m^2 m^2_{\rm pl} (1+\omega^H_{_M})}.
\label{ps1}
\end{eqnarray}
From pre-inflation $H> H_*$ to inflation $H\approx H_*$, $\rho_{_\Lambda}$ and $c_s$ 
are almost constants, and the variation $\omega^H_{_M}$ 
is $1/3$ at most. Therefore the scalar spectrum 
$\Delta^2_{_{\mathcal R}} (k)$ (\ref{ps1}) 
decreases $3/4$, as 
the scalar spectrum goes to the large distance scale of CMB observations, 
exploring the high-energy scale of horizon crossing. 
It probably explains the large-scale anomaly of
the low amplitude of the CMB power spectrum at low-$\ell$ multipole, e.g., the CMB power spectrum drops $3/4$ at $\ell=2$. These are new features of the $\tilde\Lambda$CDM scenario in the pre-inflation epoch. 
However, present discussions are preliminary, and further studies are required.
\comment{Whereas, equation (\ref{eps0}) implies that $\epsilon$ and $r\approx 16\epsilon$ increase,  
$n_s=1-2\epsilon$ decreases, due to $\omega^H_{_M}$ value increases as the pivot scale $k_*^{-1}$ 
goes to large scales.}

\section{Discussions on dark-matter density perturbations}

Since dark matter dominates over normal matter today, 
we suppose that major massive pairs produced in pre-inflation,  
inflation and reheating should be dark-matter particles. In addition to quantum pair oscillating modes (Fig.~\ref{osci}), the pair plasma oscillation appears when the massive pair plasma density $\rho^H_{_M}$ is large enough. The acoustic wave of the density perturbation $\delta \rho^H_{_M}/\rho^H_{_M}$ 
is formed and described by the sound velocity 
$c^{M}_s=(\partial p^H_{_M}/ \partial \rho^H_{_M})^{1/2}=(\omega^H_{_M})^{1/2}$. We might call these primordial modes as dark-matter density perturbations to distinguish them from curvature perturbations. 
The quantum pair oscillating modes and massive pair plasma acoustic density perturbations exited and reentered the horizon, which should imprint on both CMB and matter density power spectra. The phenomenon is similar to the usual discussions on the curvature perturbation modes imprinting on the CMB power spectra. In addition, dark-matter density perturbations in reheating exited and reentered the horizon could account for baryogenesis. Reference \cite{Xue2020b} presents the preliminary results. 
However, detailed discussions and quantitative calculations are required to see if these primordial dark-matter density perturbations are interesting modes to confront with observations. To end this article, it is worthwhile to mention that in 
Refs.~\cite{Xue2015, Begue2019, Xue2022} \footnote{The $\tilde\Lambda$CDM scenario was 
named by QFC (quantum field cosmology) in Ref.~\cite{Begue2019}.}, we study in the 
$\tilde\Lambda$CDM scenario 
the very slowly varying ``dark-energy'' $\tilde\Lambda$ interacting with radiation and matter from the reheating $\rho_{_\Lambda}\ll \rho_{_{M,R}}$ to the present time 
$\rho_{_\Lambda}>\rho_{_{M,R}}$. Reference \cite{Gao2021, Gao2022} shows their relevance 
for relieving the $H_0$ tension.
  
\section{\bf Acknowledgment}
The author thanks the Editor Alexei Starobinsky and anonymous referees for their reports that improve the article. 


\begin{thebibliography}{10}

\bibitem{PhysRevLett.21.562}
L.~Parker, \emph{Particle creation in expanding universes},
  \href{https://doi.org/10.1103/PhysRevLett.21.562}{\emph{Phys. Rev. Lett.}
  {\bfseries 21} (1968) 562}.

\bibitem{PhysRev.183.1057}
L.~Parker, \emph{Quantized fields and particle creation in expanding universes.
  I}, \href{https://doi.org/10.1103/PhysRev.183.1057}{\emph{Phys. Rev.}
  {\bfseries 183} (1969) 1057}.

\bibitem{PhysRevD.3.346}
L.~Parker, \emph{Quantized fields and particle creation in expanding universes.
  II}, \href{https://doi.org/10.1103/PhysRevD.3.346}{\emph{Phys. Rev. D}
  {\bfseries 3} (1971) 346}.

\bibitem{Zeldovich1971}
Y.~B. Zeldovich and A.~A. Starobinsky, \emph{Particle production and vacuum
  polarization in an anisotropic gravitational field}, {\emph{JETP [Zh. Eksp.
  Teor. Fiz. 61 (1971) 2161-2175]} {\bfseries 34} (1972) 1159}.

\bibitem{Birrell1984}
N.~D. Birrell and P.~C.~W. Davies, \emph{Quantum Fields in Curved Space},
  Cambridge Monographs on Mathematical Physics. Cambridge Univ. Press,
  Cambridge, UK, 2, 1984,
  \href{https://doi.org/10.1017/CBO9780511622632}{10.1017/CBO9780511622632}.

\bibitem{Mamaev1976}
V.~M.~M. S.~G.~Mamaev and A.~A. Starobinsky, \emph{Particle production and
  vacuum polarization in an anisotropic gravitational field}, {\emph{JETP}
  {\bfseries 43 (5)} (1976) 823}.

\bibitem{Zeldovich1977}
Y.~B. Zeldovich and A.~A. Starobinsky, \emph{Rate of particle production in
  gravitational fields}, {\emph{JETP Lett. 26, 252-255 (1977)} {\bfseries 26}
  (1977) 252}.

\bibitem{Starobinsky:1979ty}
A.~A. Starobinsky, \emph{Spectrum of relict gravitational radiation and the
  early state of the universe}, {\emph{JETP Lett.} {\bfseries 30} (1979) 682}.

\bibitem{Mottola1985}
E.~Mottola, \emph{Particle creation in de sitter space},
  \href{https://doi.org/10.1103/PhysRevD.31.754}{\emph{Phys. Rev. D} {\bfseries
  31} (1985) 754}.

\bibitem{Habib2000}
S.~Habib, C.~Molina-Paris and E.~Mottola, \emph{Energy momentum tensor of
  particles created in an expanding universe},
  \href{https://doi.org/10.1103/PhysRevD.61.024010}{\emph{Phys. Rev. D}
  {\bfseries 61} (2000) 024010}
  [\href{https://arxiv.org/abs/gr-qc/9906120}{{\ttfamily gr-qc/9906120}}].

\bibitem{Anderson2014}
P.~R. Anderson and E.~Mottola, \emph{Instability of global de sitter space to
  particle creation},
  \href{https://doi.org/10.1103/PhysRevD.89.104038}{\emph{Phys. Rev. D}
  {\bfseries 89} (2014) 104038}
  [\href{https://arxiv.org/abs/1310.0030}{{\ttfamily 1310.0030}}].

\bibitem{Anderson2014a}
P.~R. Anderson and E.~Mottola, \emph{Quantum vacuum instability of
  \textquotedblleft{}eternal\textquotedblright{} de sitter space},
  \href{https://doi.org/10.1103/PhysRevD.89.104039}{\emph{Phys. Rev. D}
  {\bfseries 89} (2014) 104039}
  [\href{https://arxiv.org/abs/1310.1963}{{\ttfamily 1310.1963}}].

\bibitem{Landete2014}
A.~Landete, J.~Navarro-Salas and F.~Torrenti, \emph{Adiabatic regularization
  and particle creation for spin one-half fields},
  \href{https://doi.org/10.1103/PhysRevD.89.044030}{\emph{Phys. Rev. D}
  {\bfseries 89} (2014) 044030}
  [\href{https://arxiv.org/abs/1311.4958}{{\ttfamily 1311.4958}}].

\bibitem{Parker1973}
L.~Parker and S.~A. Fulling, \emph{Quantized matter fields and the avoidance of
  singularities in general relativity},
  \href{https://doi.org/10.1103/PhysRevD.7.2357}{\emph{Phys. Rev. D} {\bfseries
  7} (1973) 2357}.

\bibitem{Starobinsky1980}
A.~A. Starobinsky, \emph{A new type of isotropic cosmological models without
  singularity}, \href{https://doi.org/10.1016/0370-2693(80)90670-X}{\emph{Phys.
  Lett. B} {\bfseries 91} (1980) 99}.

\bibitem{Starobinsky1982}
A.~A. Starobinsky, \emph{Nonsingular model of the universe with the
  quantum-gravitational de sitter stage and its observational consequences},
  {\emph{Proc. of the Second Seminar ``Quantum Theory of
  Gravity", Moscow, October 1981, INR Press, Moscow} (1982) 58}.

\bibitem{Ford1987}
L.~H. Ford, \emph{Gravitational particle creation and inflation},
  \href{https://doi.org/10.1103/PhysRevD.35.2955}{\emph{Phys. Rev. D}
  {\bfseries 35} (1987) 2955}.

\bibitem{Kolb1996}
E.~W. Kolb, A.~D. Linde and A.~Riotto, \emph{Gut baryogenesis after
  preheating}, \href{https://doi.org/10.1103/PhysRevLett.77.4290}{\emph{Phys.
  Rev. Lett.} {\bfseries 77} (1996) 4290}
  [\href{https://arxiv.org/abs/hep-ph/9606260}{{\ttfamily hep-ph/9606260}}].

\bibitem{Greene1997}
B.~R. Greene, T.~Prokopec and T.~G. Roos, \emph{Inflaton decay and heavy
  particle production with negative coupling},
  \href{https://doi.org/10.1103/PhysRevD.56.6484}{\emph{Phys. Rev. D}
  {\bfseries 56} (1997) 6484}
  [\href{https://arxiv.org/abs/hep-ph/9705357}{{\ttfamily hep-ph/9705357}}].

\bibitem{Kolb1998}
E.~W. Kolb, A.~Riotto and I.~I. Tkachev, \emph{Gut baryogenesis after
  preheating: Numerical study of the production and decay of x bosons},
  \href{https://doi.org/10.1016/S0370-2693(98)00134-8}{\emph{Phys. Lett. B}
  {\bfseries 423} (1998) 348}
  [\href{https://arxiv.org/abs/hep-ph/9801306}{{\ttfamily hep-ph/9801306}}].

\bibitem{Chung1998}
D.~J.~H. Chung, E.~W. Kolb and A.~Riotto, \emph{Superheavy dark matter},
  \href{https://doi.org/10.1103/PhysRevD.59.023501}{\emph{Phys. Rev. D}
  {\bfseries 59} (1998) 023501}
  [\href{https://arxiv.org/abs/hep-ph/9802238}{{\ttfamily hep-ph/9802238}}].

\bibitem{Chung2001}
D.~J.~H. Chung, P.~Crotty, E.~W. Kolb and A.~Riotto, \emph{On the gravitational
  production of superheavy dark matter},
  \href{https://doi.org/10.1103/PhysRevD.64.043503}{\emph{Phys. Rev. D}
  {\bfseries 64} (2001) 043503}
  [\href{https://arxiv.org/abs/hep-ph/0104100}{{\ttfamily hep-ph/0104100}}].

\bibitem{Chung2000}
D.~J.~H. Chung, E.~W. Kolb, A.~Riotto and I.~I. Tkachev, \emph{Probing
  planckian physics: Resonant production of particles during inflation and
  features in the primordial power spectrum},
  \href{https://doi.org/10.1103/PhysRevD.62.043508}{\emph{Phys. Rev. D}
  {\bfseries 62} (2000) 043508}
  [\href{https://arxiv.org/abs/hep-ph/9910437}{{\ttfamily hep-ph/9910437}}].

\bibitem{Chung2003}
D.~J.~H. Chung, \emph{Classical inflation field induced creation of superheavy
  dark matter}, \href{https://doi.org/10.1103/PhysRevD.67.083514}{\emph{Phys.
  Rev. D} {\bfseries 67} (2003) 083514}
  [\href{https://arxiv.org/abs/hep-ph/9809489}{{\ttfamily hep-ph/9809489}}].

\bibitem{Chung2005}
D.~J.~H. Chung, E.~W. Kolb, A.~Riotto and L.~Senatore, \emph{Isocurvature
  constraints on gravitationally produced superheavy dark matter},
  \href{https://doi.org/10.1103/PhysRevD.72.023511}{\emph{Phys. Rev. D}
  {\bfseries 72} (2005) 023511}
  [\href{https://arxiv.org/abs/astro-ph/0411468}{{\ttfamily
  astro-ph/0411468}}].

\bibitem{Ema2016}
Y.~Ema, R.~Jinno, K.~Mukaida and K.~Nakayama, \emph{Gravitational particle
  production in oscillating backgrounds and its cosmological implications},
  \href{https://doi.org/10.1103/PhysRevD.94.063517}{\emph{Phys. Rev. D}
  {\bfseries 94} (2016) 063517}
  [\href{https://arxiv.org/abs/1604.08898}{{\ttfamily 1604.08898}}].

\bibitem{Chung2019}
D.~J.~H. Chung, E.~W. Kolb and A.~J. Long, \emph{Gravitational production of
  super-hubble-mass particles: an analytic approach},
  \href{https://doi.org/10.1007/JHEP01(2019)189}{\emph{JHEP} {\bfseries 01}
  (2019) 189} [\href{https://arxiv.org/abs/1812.00211}{{\ttfamily
  1812.00211}}].

\bibitem{Ema2018}
Y.~Ema, K.~Nakayama and Y.~Tang, \emph{Production of purely gravitational dark
  matter}, \href{https://doi.org/10.1007/JHEP09(2018)135}{\emph{JHEP}
  {\bfseries 09} (2018) 135}
  [\href{https://arxiv.org/abs/1804.07471}{{\ttfamily 1804.07471}}].

\bibitem{Li2019}
L.~Li, T.~Nakama, C.~M. Sou, Y.~Wang and S.~Zhou, \emph{Gravitational
  production of superheavy dark matter and associated cosmological signatures},
  \href{https://doi.org/10.1007/JHEP07(2019)067}{\emph{JHEP} {\bfseries 07}
  (2019) 067} [\href{https://arxiv.org/abs/1903.08842}{{\ttfamily
  1903.08842}}].

\bibitem{Xue2019}
S.-S. Xue, \emph{Cosmological $\Lambda$ driven inflation and produced massive
  particles},  \href{https://arxiv.org/abs/1910.03938}{{\ttfamily 1910.03938}}.

\bibitem{Xue2020}
S.-S. Xue, \emph{Cosmological constant, matter, cosmic inflation and
  coincidence}, \href{https://doi.org/10.1142/S0217732320501230}{\emph{Mod.
  Phys. Lett. A} {\bfseries 35} (2020) 2050123}
  [\href{https://arxiv.org/abs/2004.10859}{{\ttfamily 2004.10859}}].

\bibitem{Kofman1994}
L.~Kofman, A.~D. Linde and A.~A. Starobinsky, \emph{Reheating after inflation},
  \href{https://doi.org/10.1103/PhysRevLett.73.3195}{\emph{Phys. Rev. Lett.}
  {\bfseries 73} (1994) 3195}
  [\href{https://arxiv.org/abs/hep-th/9405187}{{\ttfamily hep-th/9405187}}].

\bibitem{Kofman1997}
L.~Kofman, A.~D. Linde and A.~A. Starobinsky, \emph{Towards the theory of
  reheating after inflation},
  \href{https://doi.org/10.1103/PhysRevD.56.3258}{\emph{Phys. Rev. D}
  {\bfseries 56} (1997) 3258}
  [\href{https://arxiv.org/abs/hep-ph/9704452}{{\ttfamily hep-ph/9704452}}].

\bibitem{Kuzmin1998}
V.~Kuzmin and I.~Tkachev, \emph{Ultrahigh-energy cosmic rays, superheavy long
  living particles, and matter creation after inflation},
  \href{https://doi.org/10.1134/1.567858}{\emph{JETP Lett.} {\bfseries 68}
  (1998) 271} [\href{https://arxiv.org/abs/hep-ph/9802304}{{\ttfamily
  hep-ph/9802304}}].

\bibitem{Kuzmin1999}
V.~Kuzmin and I.~Tkachev, \emph{Matter creation via vacuum fluctuations in the
  early universe and observed ultrahigh-energy cosmic ray events},
  \href{https://doi.org/10.1103/PhysRevD.59.123006}{\emph{Phys. Rev. D}
  {\bfseries 59} (1999) 123006}
  [\href{https://arxiv.org/abs/hep-ph/9809547}{{\ttfamily hep-ph/9809547}}].

\bibitem{Kolb1999}
E.~W. Kolb, D.~J.~H. Chung and A.~Riotto, \emph{Wimpzillas!},
  \href{https://doi.org/10.1063/1.59655}{\emph{AIP Conf. Proc.} {\bfseries 484}
  (1999) 91} [\href{https://arxiv.org/abs/hep-ph/9810361}{{\ttfamily
  hep-ph/9810361}}].

\bibitem{Bertone2005}
G.~Bertone, D.~Hooper and J.~Silk, \emph{Particle dark matter: Evidence,
  candidates and constraints},
  \href{https://doi.org/10.1016/j.physrep.2004.08.031}{\emph{Phys. Rept.}
  {\bfseries 405} (2005) 279}
  [\href{https://arxiv.org/abs/hep-ph/0404175}{{\ttfamily hep-ph/0404175}}].

\bibitem{Bassett2006}
B.~A. Bassett, S.~Tsujikawa and D.~Wands, \emph{Inflation dynamics and
  reheating}, \href{https://doi.org/10.1103/RevModPhys.78.537}{\emph{Rev. Mod.
  Phys.} {\bfseries 78} (2006) 537}
  [\href{https://arxiv.org/abs/astro-ph/0507632}{{\ttfamily
  astro-ph/0507632}}].

\bibitem{Kolb2007}
E.~W. Kolb, A.~A. Starobinsky and I.~I. Tkachev, \emph{Trans-planckian
  wimpzillas}, \href{https://doi.org/10.1088/1475-7516/2007/07/005}{\emph{JCAP}
  {\bfseries 07} (2007) 005}
  [\href{https://arxiv.org/abs/hep-th/0702143}{{\ttfamily hep-th/0702143}}].

\bibitem{Allahverdi2010}
R.~Allahverdi, R.~Brandenberger, F.-Y. Cyr-Racine and A.~Mazumdar,
  \emph{Reheating in inflationary cosmology: Theory and applications},
  \href{https://doi.org/10.1146/annurev.nucl.012809.104511}{\emph{Ann. Rev.
  Nucl. Part. Sci.} {\bfseries 60} (2010) 27}
  [\href{https://arxiv.org/abs/1001.2600}{{\ttfamily 1001.2600}}].

\bibitem{Frolov2010}
A.~V. Frolov, \emph{Non-linear dynamics and primordial curvature perturbations
  from preheating},
  \href{https://doi.org/10.1088/0264-9381/27/12/124006}{\emph{Class. Quant.
  Grav.} {\bfseries 27} (2010) 124006}
  [\href{https://arxiv.org/abs/1004.3559}{{\ttfamily 1004.3559}}].

\bibitem{Hall2010}
L.~J. Hall, K.~Jedamzik, J.~March-Russell and S.~M. West, \emph{Freeze-in
  production of fimp dark matter},
  \href{https://doi.org/10.1007/JHEP03(2010)080}{\emph{JHEP} {\bfseries 03}
  (2010) 080} [\href{https://arxiv.org/abs/0911.1120}{{\ttfamily 0911.1120}}].

\bibitem{Feng2010}
J.~L. Feng, \emph{Dark matter candidates from particle physics and methods of
  detection},
  \href{https://doi.org/10.1146/annurev-astro-082708-101659}{\emph{Ann. Rev.
  Astron. Astrophys.} {\bfseries 48} (2010) 495}
  [\href{https://arxiv.org/abs/1003.0904}{{\ttfamily 1003.0904}}].

\bibitem{Gorbunov2012}
D.~S. Gorbunov and A.~G. Panin, \emph{Free scalar dark matter candidates in
  $R^2$-inflation: the light, the heavy and the superheavy},
  \href{https://doi.org/10.1016/j.physletb.2012.10.015}{\emph{Phys. Lett. B}
  {\bfseries 718} (2012) 15} [\href{https://arxiv.org/abs/1201.3539}{{\ttfamily
  1201.3539}}].

\bibitem{Amin2014}
M.~A. Amin, M.~P. Hertzberg, D.~I. Kaiser and J.~Karouby, \emph{Nonperturbative
  dynamics of reheating after inflation: A review},
  \href{https://doi.org/10.1142/S0218271815300037}{\emph{Int. J. Mod. Phys. D}
  {\bfseries 24} (2014) 1530003}
  [\href{https://arxiv.org/abs/1410.3808}{{\ttfamily 1410.3808}}].

\bibitem{Ema2015}
Y.~Ema, R.~Jinno, K.~Mukaida and K.~Nakayama, \emph{Gravitational effects on
  inflaton decay},
  \href{https://doi.org/10.1088/1475-7516/2015/05/038}{\emph{JCAP} {\bfseries
  05} (2015) 038} [\href{https://arxiv.org/abs/1502.02475}{{\ttfamily
  1502.02475}}].

\bibitem{Garny2016}
M.~Garny, M.~Sandora and M.~S. Sloth, \emph{Planckian interacting massive
  particles as dark matter},
  \href{https://doi.org/10.1103/PhysRevLett.116.101302}{\emph{Phys. Rev. Lett.}
  {\bfseries 116} (2016) 101302}
  [\href{https://arxiv.org/abs/1511.03278}{{\ttfamily 1511.03278}}].

\bibitem{Garny2018}
M.~Garny, A.~Palessandro, M.~Sandora and M.~S. Sloth, \emph{Theory and
  phenomenology of planckian interacting massive particles as dark matter},
  \href{https://doi.org/10.1088/1475-7516/2018/02/027}{\emph{JCAP} {\bfseries
  02} (2018) 027} [\href{https://arxiv.org/abs/1709.09688}{{\ttfamily
  1709.09688}}].

\bibitem{Kolb2017}
E.~W. Kolb and A.~J. Long, \emph{Superheavy dark matter through higgs portal
  operators}, \href{https://doi.org/10.1103/PhysRevD.96.103540}{\emph{Phys.
  Rev. D} {\bfseries 96} (2017) 103540}
  [\href{https://arxiv.org/abs/1708.04293}{{\ttfamily 1708.04293}}].

\bibitem{Hashiba2019}
S.~Hashiba and J.~Yokoyama, \emph{Gravitational reheating through conformally
  coupled superheavy scalar particles},
  \href{https://doi.org/10.1088/1475-7516/2019/01/028}{\emph{JCAP} {\bfseries
  01} (2019) 028} [\href{https://arxiv.org/abs/1809.05410}{{\ttfamily
  1809.05410}}].

\bibitem{Hashiba2019a}
S.~Hashiba and J.~Yokoyama, \emph{Gravitational particle creation for dark
  matter and reheating},
  \href{https://doi.org/10.1103/PhysRevD.99.043008}{\emph{Phys. Rev. D}
  {\bfseries 99} (2019) 043008}
  [\href{https://arxiv.org/abs/1812.10032}{{\ttfamily 1812.10032}}].

\bibitem{Haro2019}
J.~Haro, W.~Yang and S.~Pan, \emph{Reheating in quintessential inflation via
  gravitational production of heavy massive particles: A detailed analysis},
  \href{https://doi.org/10.1088/1475-7516/2019/01/023}{\emph{JCAP} {\bfseries
  01} (2019) 023} [\href{https://arxiv.org/abs/1811.07371}{{\ttfamily
  1811.07371}}].

\bibitem{Xue2020a}
S.-S. Xue, \emph{Cosmological $\Lambda$ converts to reheating energy and cold
  dark matter},  \href{https://arxiv.org/abs/2006.15622}{{\ttfamily
  2006.15622}}.

\bibitem{Xue2015}
S.-S. Xue, \emph{How universe evolves with cosmological and gravitational
  constants},
  \href{https://doi.org/10.1016/j.nuclphysb.2015.05.022}{\emph{Nucl. Phys. B}
  {\bfseries 897} (2015) 326}
  [\href{https://arxiv.org/abs/1410.6152}{{\ttfamily 1410.6152}}].

\bibitem{Linde:1981mu}
A.~D. Linde, \emph{A new inflationary universe scenario: A possible solution of
  the horizon, flatness, homogeneity, isotropy and primordial monopole
  problems}, \href{https://doi.org/10.1016/0370-2693(82)91219-9}{\emph{Phys.
  Lett. B} {\bfseries 108} (1982) 389–393}.

\bibitem{Mukhanov:1982nu}
V.~F. Mukhanov and G.~V. Chibisov, \emph{The vacuum energy and large scale
  structure of the universe}, {\emph{Sov. Phys. JETP} {\bfseries 56} (1982)
  258–265}.

\bibitem{Kallosh:2021mnu}
R.~Kallosh and A.~Linde, \emph{Bicep/keck and cosmological attractors},
  \href{https://doi.org/10.1088/1475- 7516/2021/12/008}{\emph{JCAP} {\bfseries
  12} (2021) 008} [\href{https://arxiv.org/abs/2110.10902}{{\ttfamily
  2110.10902}}].

\bibitem{Parker1974}
L.~Parker and S.~A. Fulling, \emph{Adiabatic regularization of the energy
  momentum tensor of a quantized field in homogeneous spaces},
  \href{https://doi.org/10.1103/PhysRevD.9.341}{\emph{Phys. Rev. D} {\bfseries
  9} (1974) 341}.

\bibitem{Kluger1991}
Y.~Kluger, J.~M. Eisenberg, B.~Svetitsky, F.~Cooper and E.~Mottola, \emph{Pair
  production in a strong electric field},
  \href{https://doi.org/10.1103/PhysRevLett.67.2427}{\emph{Phys. Rev. Lett.}
  {\bfseries 67} (1991) 2427}.

\bibitem{Ruffini2010}
R.~Ruffini, G.~Vereshchagin and S.-S. Xue, \emph{Electron-positron pairs in
  physics and astrophysics: from heavy nuclei to black holes},
  \href{https://doi.org/10.1016/j.physrep.2009.10.004}{\emph{Phys. Rept.}
  {\bfseries 487} (2010) 1} [\href{https://arxiv.org/abs/0910.0974}{{\ttfamily
  0910.0974}}].

\bibitem{Starobinsky1978}
A.~A. Starobinsky, \emph{On one nonsingular isotropic cosmological model},
  {\emph{Sov. Astron. Lett.} {\bfseries 4} (1978) 82}.

\bibitem{Xue2010}
S.-S. Xue, \emph{Detailed discussions and calculations of quantum Regge
  calculus of Einstein-Cartan theory},
  \href{https://doi.org/10.1103/PhysRevD.82.064039}{\emph{Phys. Rev. D}
  {\bfseries 82} (2010) 064039}
  [\href{https://arxiv.org/abs/0912.2435}{{\ttfamily 0912.2435}}].

\bibitem{Xue2012}
S.-S. Xue, \emph{The phase and critical point of quantum Einstein-Cartan
  gravity}, \href{https://doi.org/10.1016/j.physletb.2012.04.024}{\emph{Phys.
  Lett. B} {\bfseries 711} (2012) 404}
  [\href{https://arxiv.org/abs/1112.1323}{{\ttfamily 1112.1323}}].

\bibitem{Xue2009}
S.-S. Xue, \emph{Quantum Regge calculus of Einstein-Cartan theory},
  \href{https://doi.org/10.1016/j.physletb.2009.10.082}{\emph{Phys. Lett. B}
  {\bfseries 682} (2009) 300}
  [\href{https://arxiv.org/abs/0902.3407}{{\ttfamily 0902.3407}}].

\bibitem{Baumann2015}
D.~Baumann and L.~McAllister, \emph{Inflation and String Theory}, Cambridge
  Monographs on Mathematical Physics. Cambridge University Press, 5, 2015,
  \href{https://doi.org/10.1017/CBO9781316105733}{10.1017/CBO9781316105733},
  [\href{https://arxiv.org/abs/1404.2601}{{\ttfamily 1404.2601}}].

\bibitem{Aghanim2020}
{\scshape Planck} collaboration, \emph{Planck 2018 results. VI. cosmological
  parameters}, \href{https://doi.org/10.1051/0004-6361/201833910}{\emph{Astron.
  Astrophys.} {\bfseries 641} (2020) A6}
  [\href{https://arxiv.org/abs/1807.06209}{{\ttfamily 1807.06209}}].

\bibitem{Ade2021}
{\scshape BICEP, Keck} collaboration, \emph{Improved constraints on primordial
  gravitational waves using Planck, WMAP, and BICEP/Keck observations through
  the 2018 observing season},
  \href{https://doi.org/10.1103/PhysRevLett.127.151301}{\emph{Phys. Rev. Lett.}
  {\bfseries 127} (2021) 151301}
  [\href{https://arxiv.org/abs/2110.00483}{{\ttfamily 2110.00483}}].

\bibitem{Tristram2021}
M.~Tristram et~al., \emph{Planck constraints on the tensor-to-scalar ratio},
  \href{https://doi.org/10.1051/0004-6361/202039585}{\emph{Astron. Astrophys.}
  {\bfseries 647} (2021) A128}
  [\href{https://arxiv.org/abs/2010.01139}{{\ttfamily 2010.01139}}].

\bibitem{Akrami2020}
{\scshape Planck} collaboration, \emph{Planck 2018 results. X. constraints on
  inflation}, \href{https://doi.org/10.1051/0004-6361/201833887}{\emph{Astron.
  Astrophys.} {\bfseries 641} (2020) A10}
  [\href{https://arxiv.org/abs/1807.06211}{{\ttfamily 1807.06211}}].

\bibitem{Weinberg2010}
S.~Weinberg, \emph{Asymptotically safe inflation},
  \href{https://doi.org/10.1103/PhysRevD.81.083535}{\emph{Phys. Rev. D}
  {\bfseries 81} (2010) 083535}
  [\href{https://arxiv.org/abs/0911.3165}{{\ttfamily 0911.3165}}].

\bibitem{Starobinsky1998}
A.~A. Starobinsky, \emph{How to determine an effective potential for a variable
  cosmological term}, \href{https://doi.org/10.1134/1.567941}{\emph{JETP Lett.
  68 (1998) 757-763; Pisma Zh.Eksp.Teor.Fiz. 68 (1998) 721-726} (1998) }
  [\href{https://arxiv.org/abs/astro-ph/9810431}{{\ttfamily
  astro-ph/9810431}}].

\bibitem{Koshelev2016}
A.~S. Koshelev, L.~Modesto, L.~Rachwal and A.~A. Starobinsky, \emph{Occurrence
  of exact $R^2$ inflation in non-local uv-complete gravity},
  \href{https://doi.org/10.1007/JHEP11(2016)067}{\emph{JHEP} {\bfseries 11}
  (2016) 067} [\href{https://arxiv.org/abs/1604.03127}{{\ttfamily
  1604.03127}}].

\bibitem{Xue2020b}
S.-S. Xue, \emph{Horizon crossing causes baryogenesis, magnetogenesis and
  dark-matter acoustic wave},
  \href{https://arxiv.org/abs/2007.03464}{{\ttfamily 2007.03464}}.

\bibitem{Begue2019}
D.~B\'egu\'e, C.~Stahl and S.-S. Xue, \emph{A model of interacting dark fluids
  tested with supernovae and baryon acoustic oscillations data},
  \href{https://doi.org/10.1016/j.nuclphysb.2019.01.001}{\emph{Nucl. Phys. B}
  {\bfseries 940} (2019) 312}
  [\href{https://arxiv.org/abs/1702.03185}{{\ttfamily 1702.03185}}].

\bibitem{Xue2022}
S.-S. Xue, \emph{Massive particle pair production and oscillation in Friedman
  universe: dark energy and matter interaction},
  \href{https://arxiv.org/abs/2203.11918}{{\ttfamily 2203.11918}}.

\bibitem{Gao2021}
L.-Y. Gao, Z.-W. Zhao, S.-S. Xue and X.~Zhang, \emph{Relieving the $H_0$ tension
  with a new interacting dark energy model},
  \href{https://doi.org/10.1088/1475-7516/2021/07/005}{\emph{JCAP} {\bfseries
  07} (2021) 005} [\href{https://arxiv.org/abs/2101.10714}{{\ttfamily
  2101.10714}}].

\bibitem{Gao2022}
L.-Y. Gao, S.-S. Xue and X.~Zhang, \emph{Dark energy and matter interacting scenario can relieve $H_0$ and $S_8$ tensions},
   \href{https://arxiv.org/abs/2212.13146}{{\ttfamily 2212.13146}}.

\end{thebibliography}

\providecommand{\href}[2]{#2}\begingroup\raggedright\endgroup

\newpage

\section{\bf Supplemental Material: quantum pair oscillation details}\label{dis}
In microscopic time, we plot the Bogoliubov coefficient $|\beta|^2$, the quantum pair density $\rho^{\rm fast}_{_\Lambda}$ and pressure $p^{\rm fast}_{_\Lambda}$, as well as the fast components of Hubble function $H_{\rm fast}$ 
and cosmological term $\rho^{\rm fast}_{_\Lambda}$. 

\begin{figure*}[h]
\vspace{+3em}
\includegraphics[height=5.5cm,width=7.8cm]{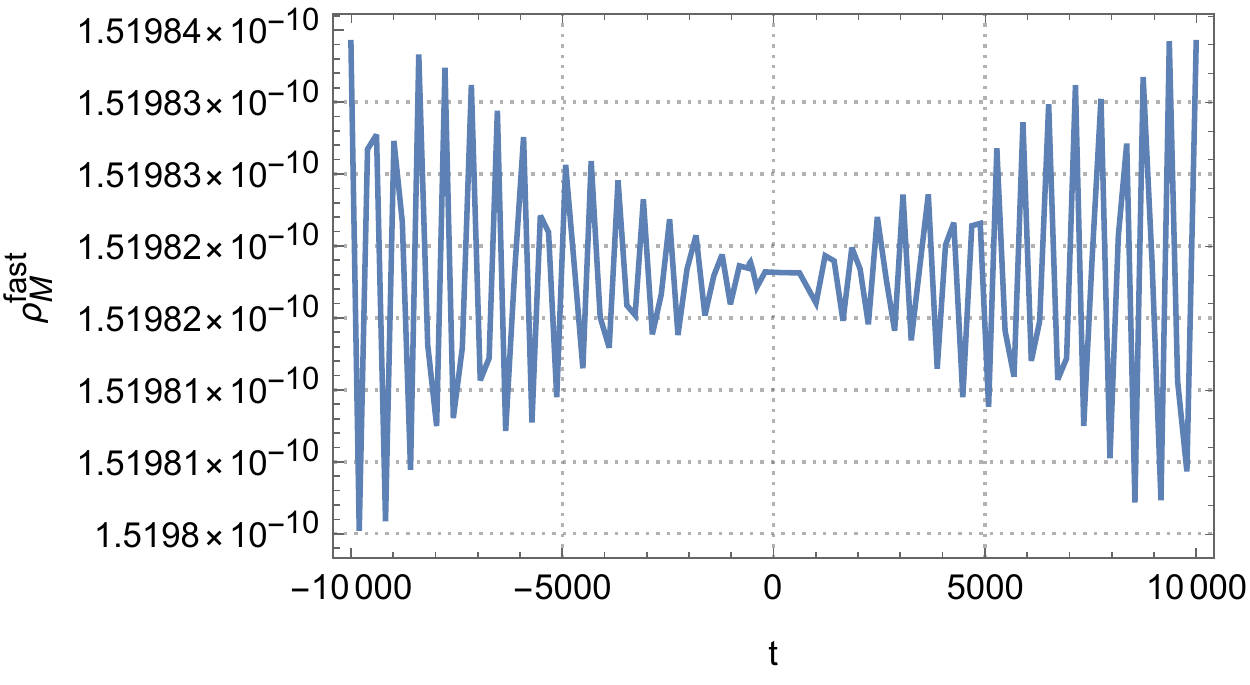}\hspace{0.333cm}
\includegraphics[height=5.5cm,width=7.8cm]{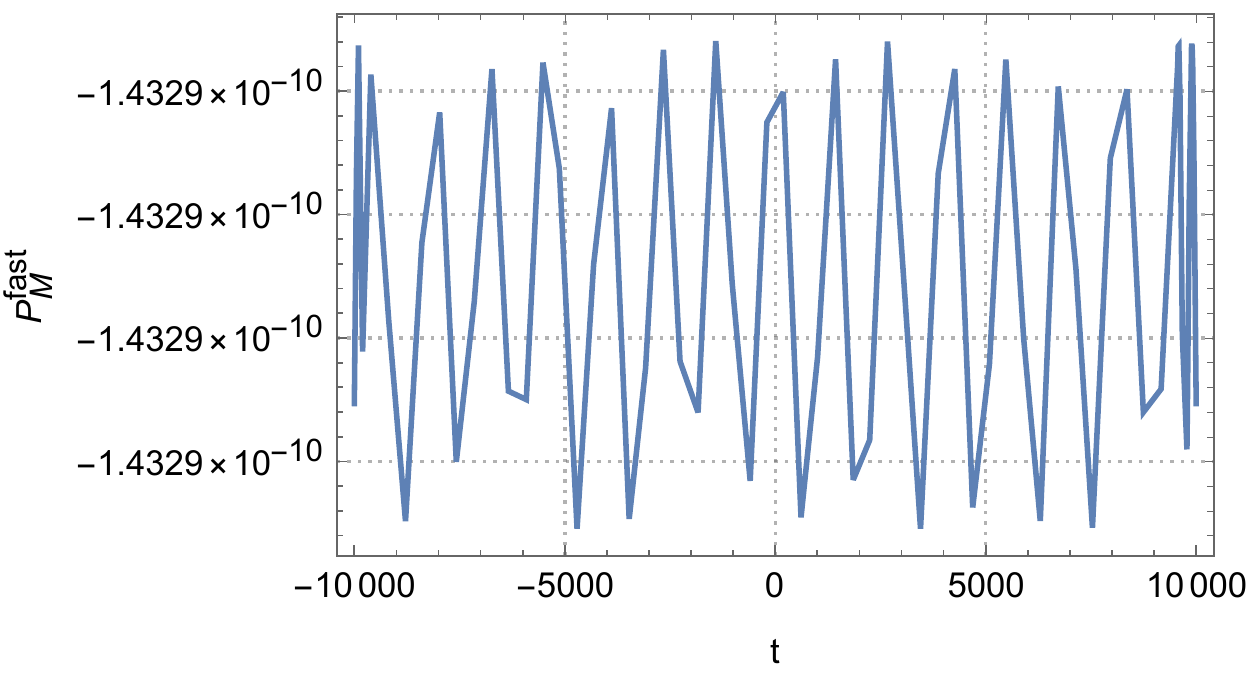}\hspace{0.333cm}
\includegraphics[height=5.5cm,width=7.8cm]{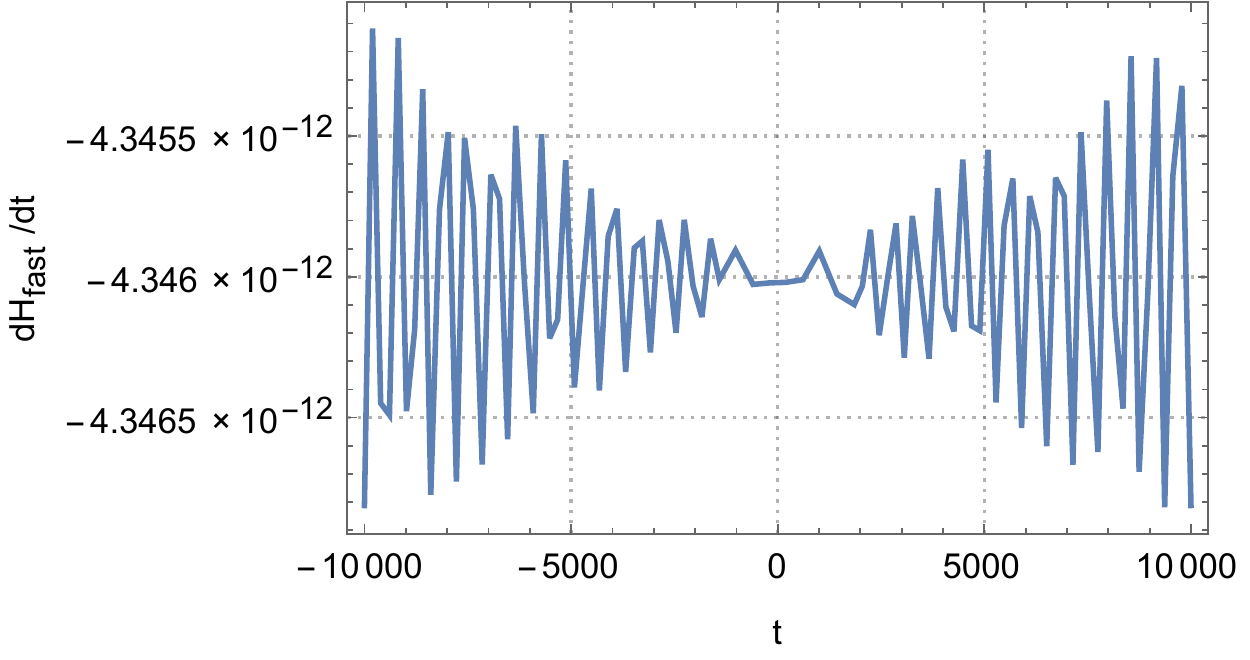}\hspace{0.333cm}
\includegraphics[height=5.5cm,width=7.8cm]{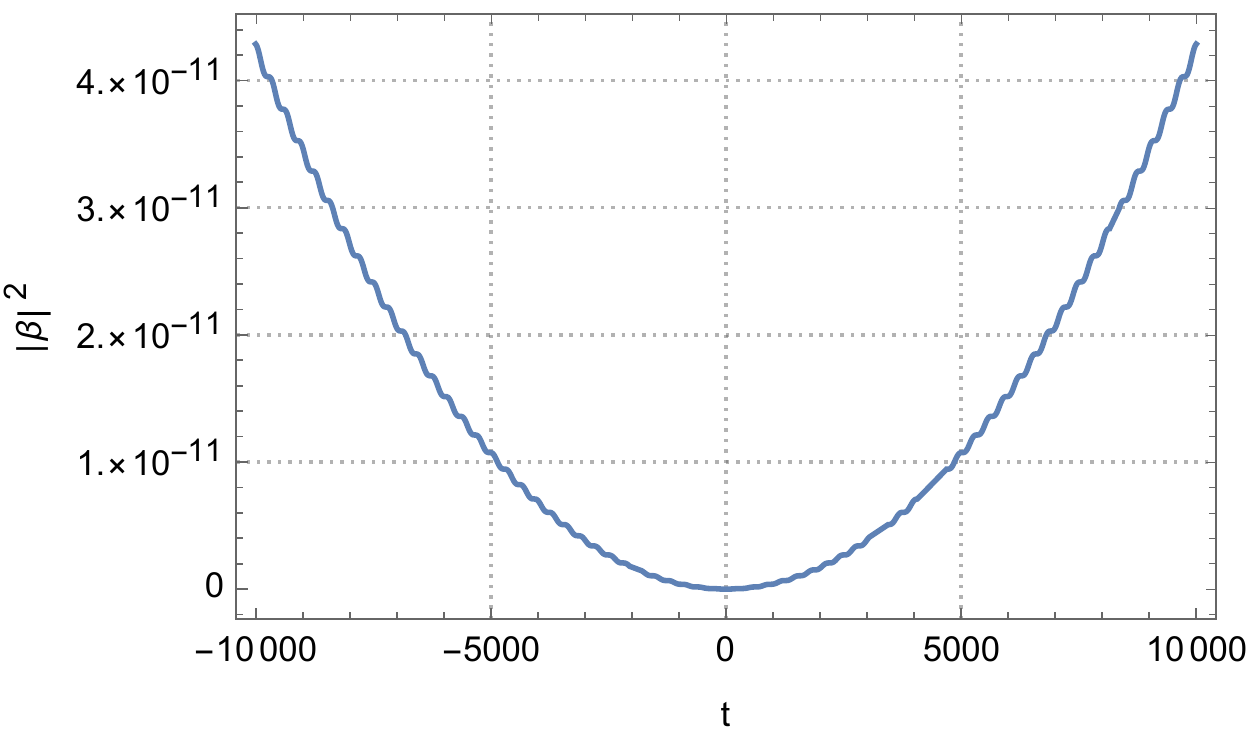}
\includegraphics[height=5.5cm,width=7.8cm]{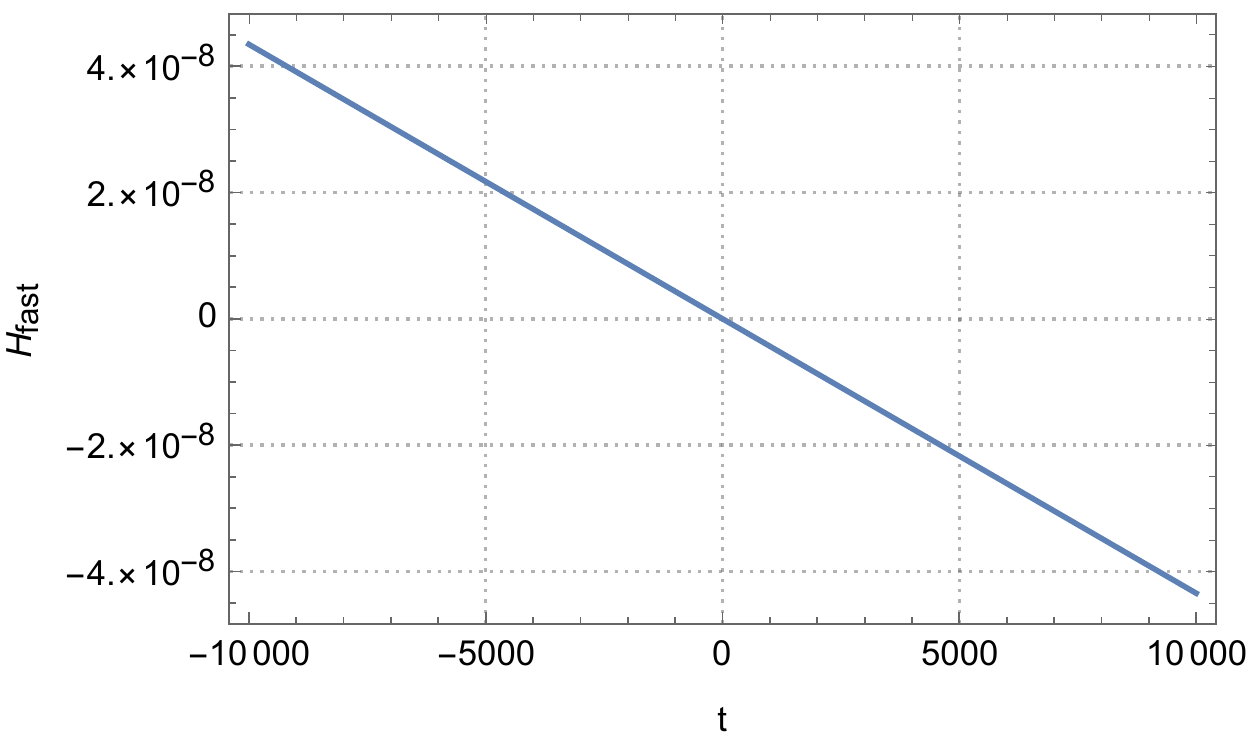}\hspace{0.333cm}
\includegraphics[height=5.5cm,width=7.8cm]{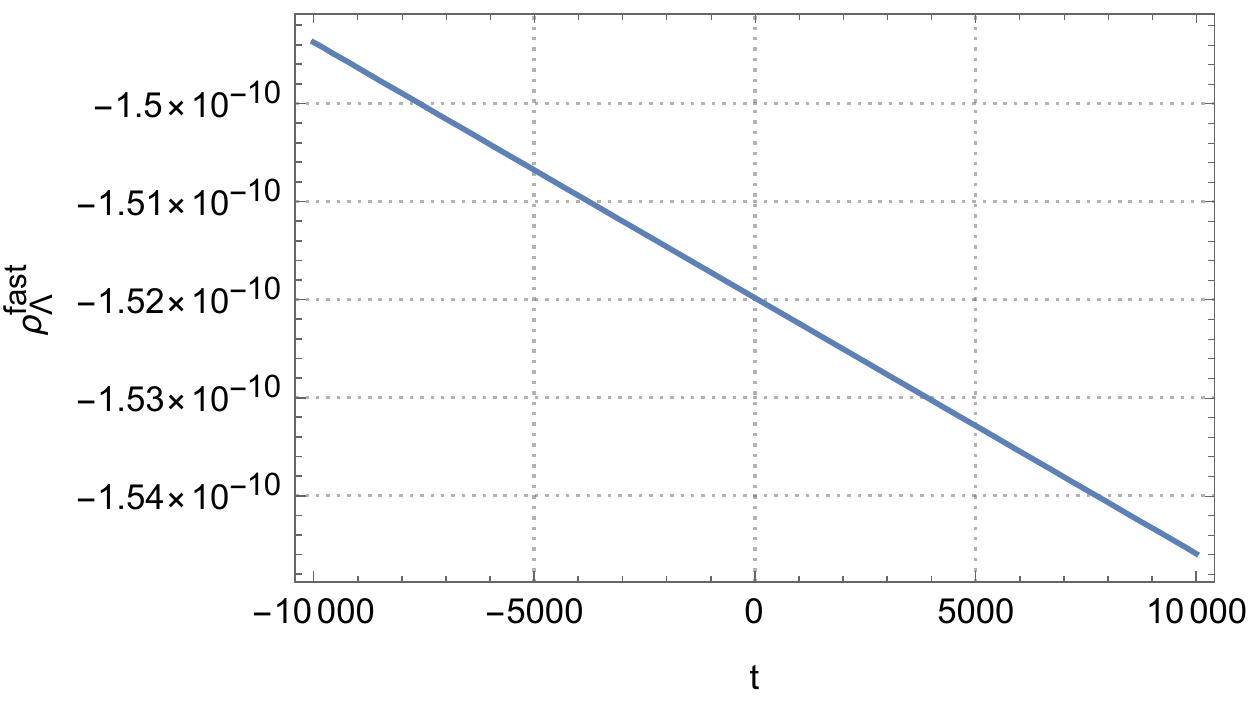}
\caption{Corresponding to Fig.~\ref{osci}, the details of quantum pair oscillation 
in a tiny layer $\lambda_MH^{-2}_{\rm slow}$ are shown. The oscillatory $H_{\rm fast}$ and $\rho^{\rm fast}_{_\Lambda}$ structures are too small to see. 
}\label{detailosci1}
\end{figure*}

\newpage

\begin{figure*}[t]
\vspace{-3em}
\includegraphics[height=5.5cm,width=7.8cm]{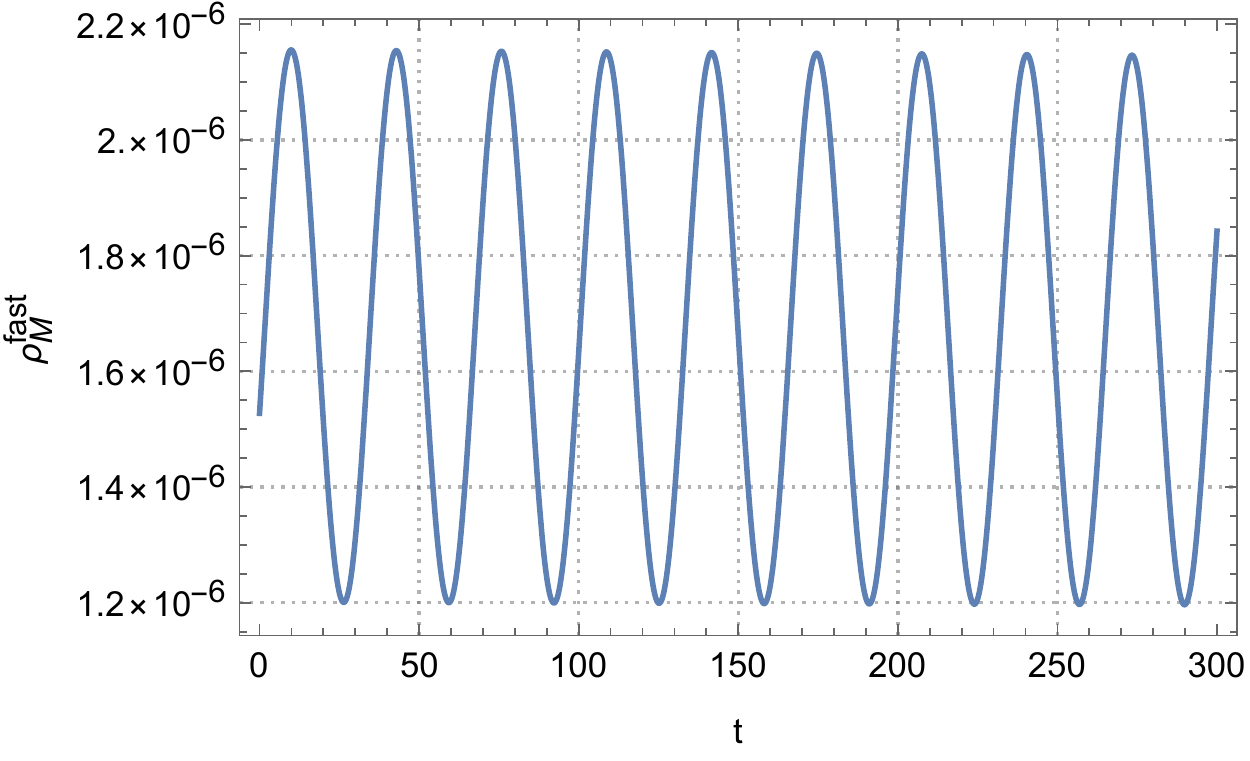}\hspace{0.333cm}
\includegraphics[height=5.5cm,width=7.8cm]{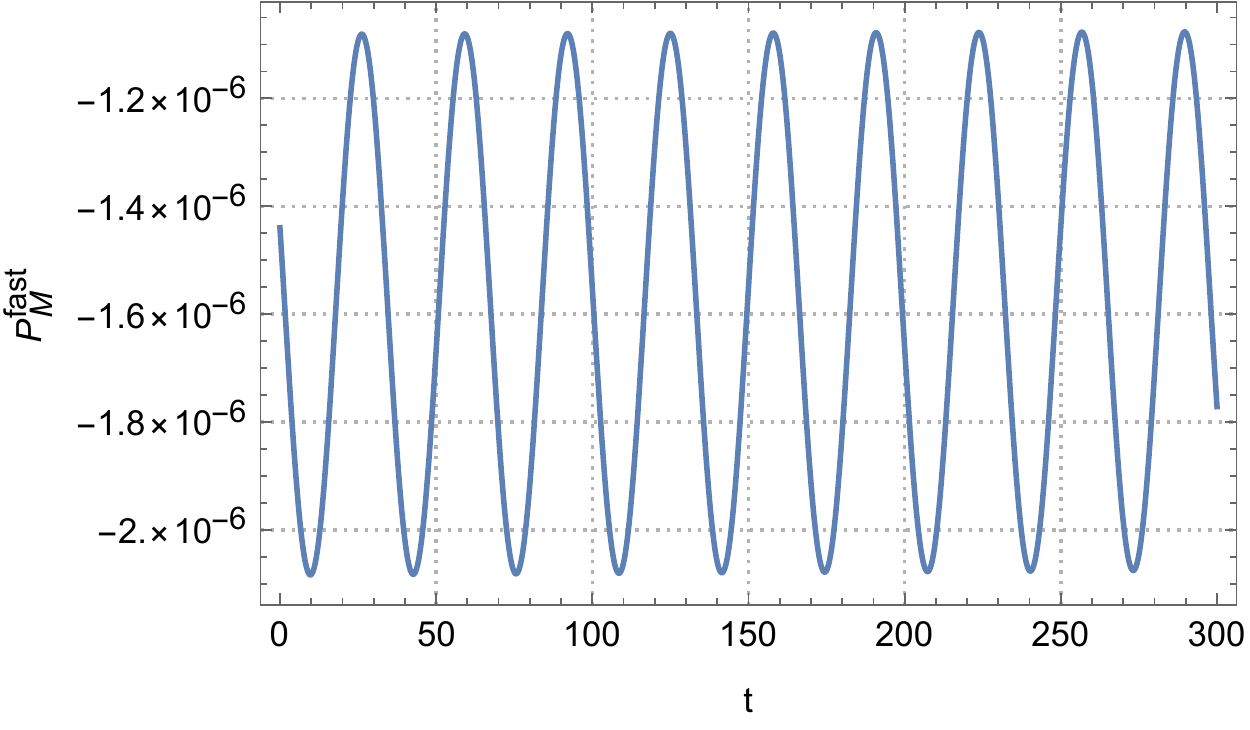}\hspace{0.333cm}
\includegraphics[height=5.5cm,width=7.8cm]{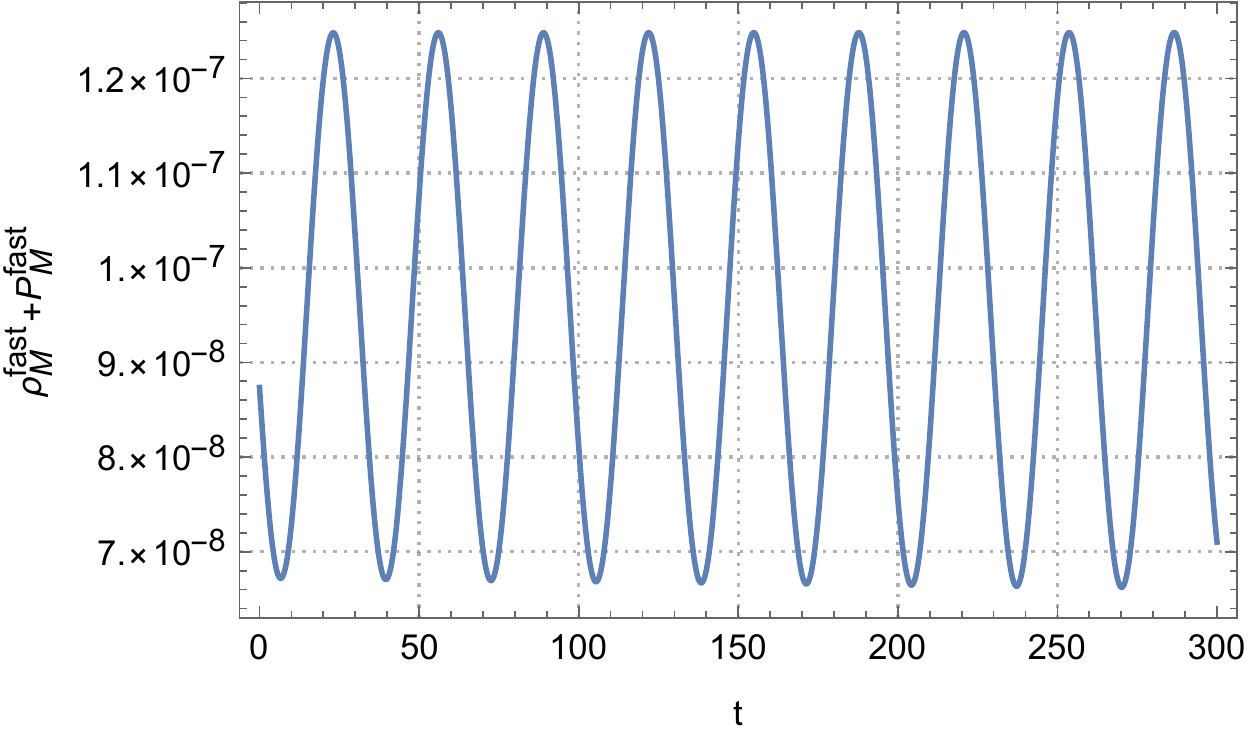}\hspace{0.333cm}
\includegraphics[height=5.5cm,width=7.8cm]{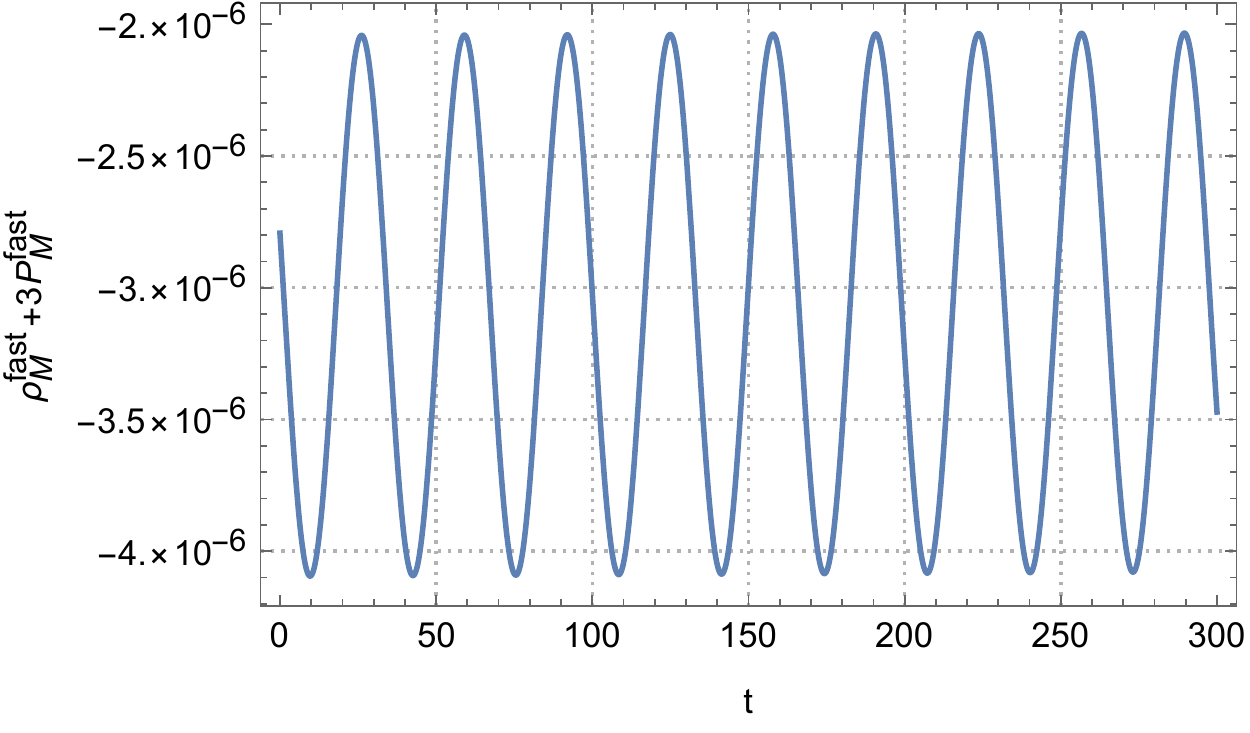}\hspace{0.333cm}
\includegraphics[height=5.5cm,width=7.8cm]{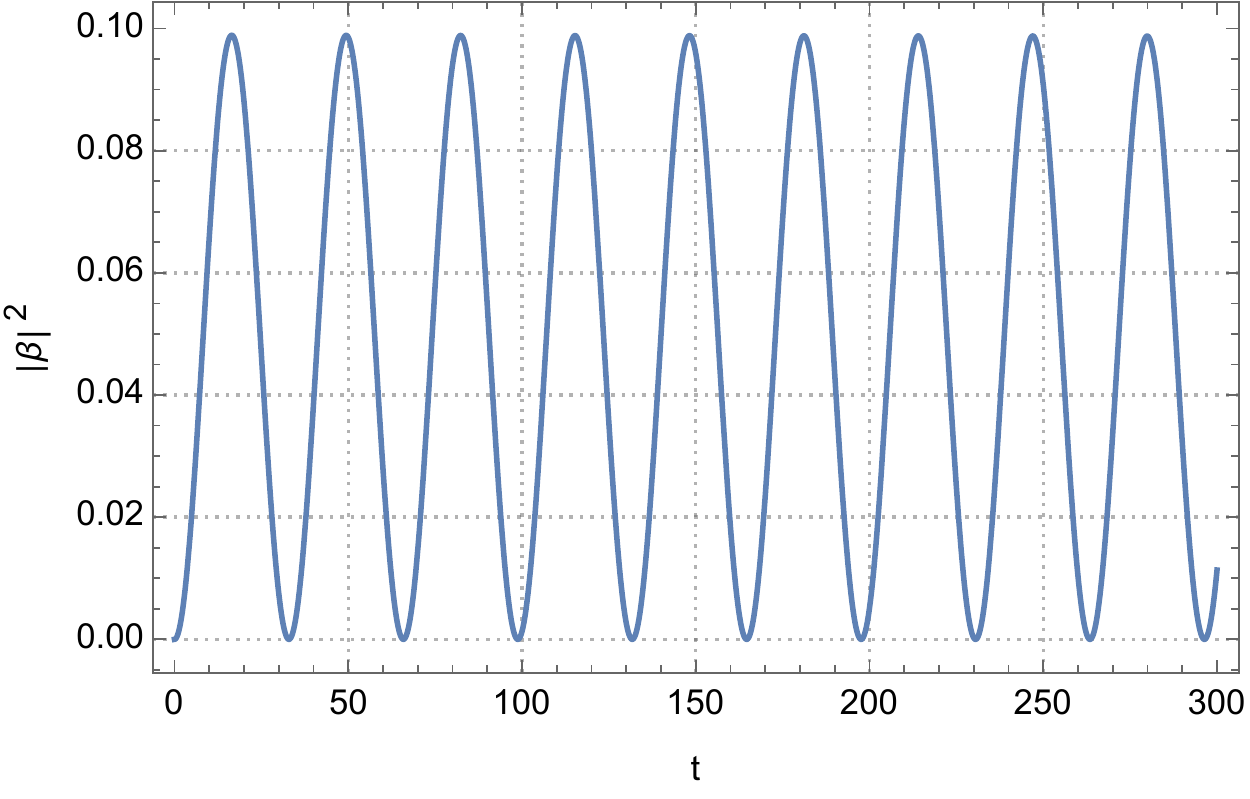}
\includegraphics[height=5.5cm,width=7.8cm]{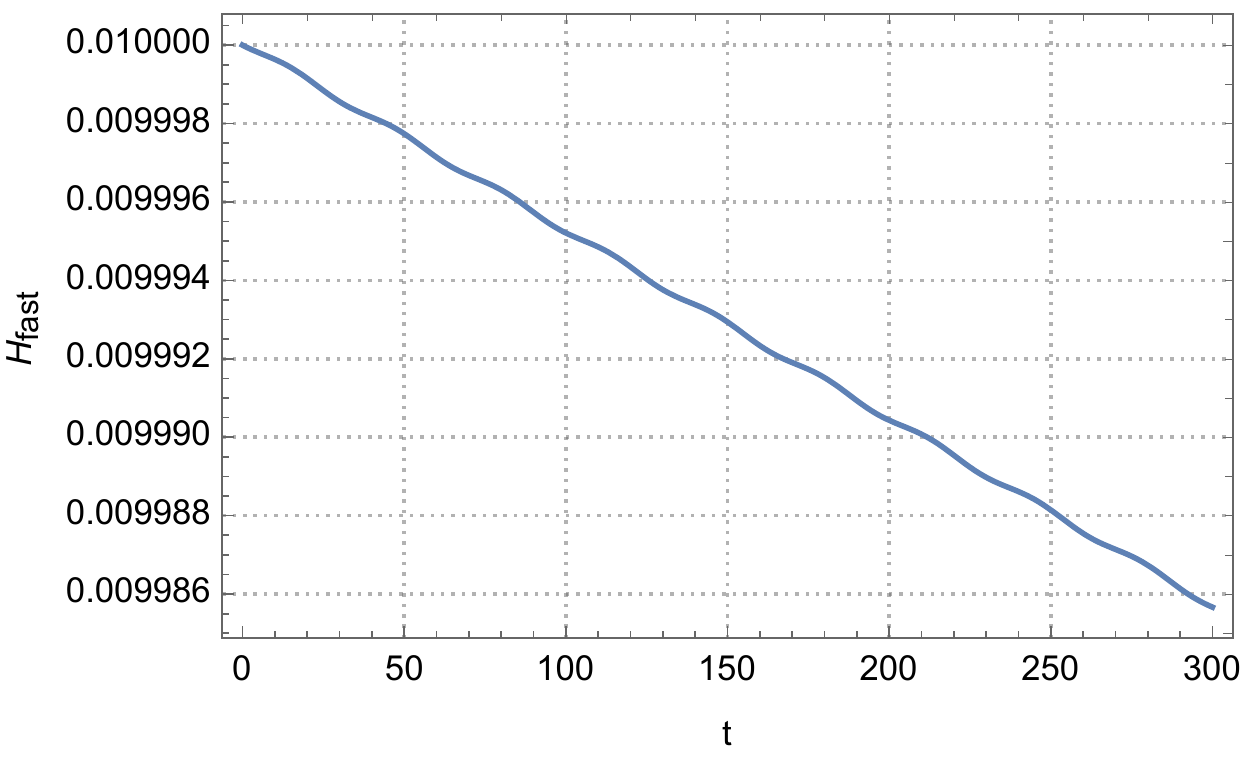}\hspace{0.333cm}
\includegraphics[height=5.5cm,width=7.8cm]{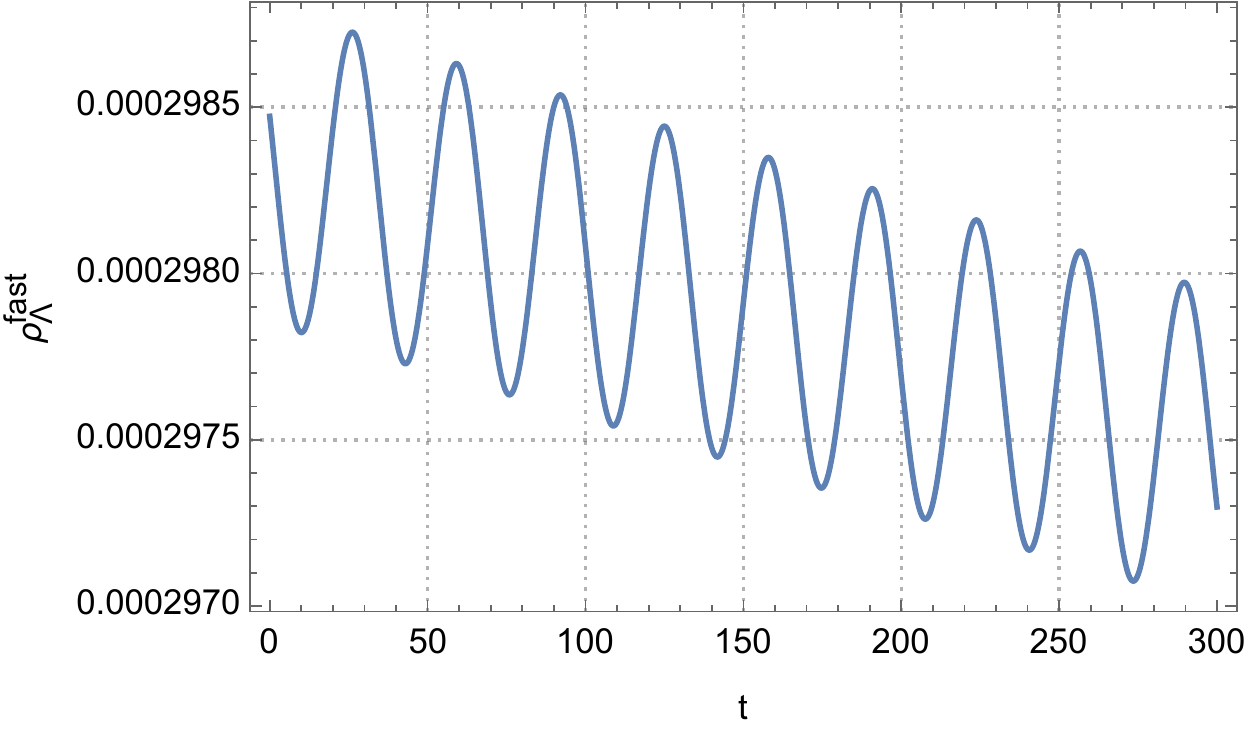}
\includegraphics[height=5.5cm,width=7.8cm]{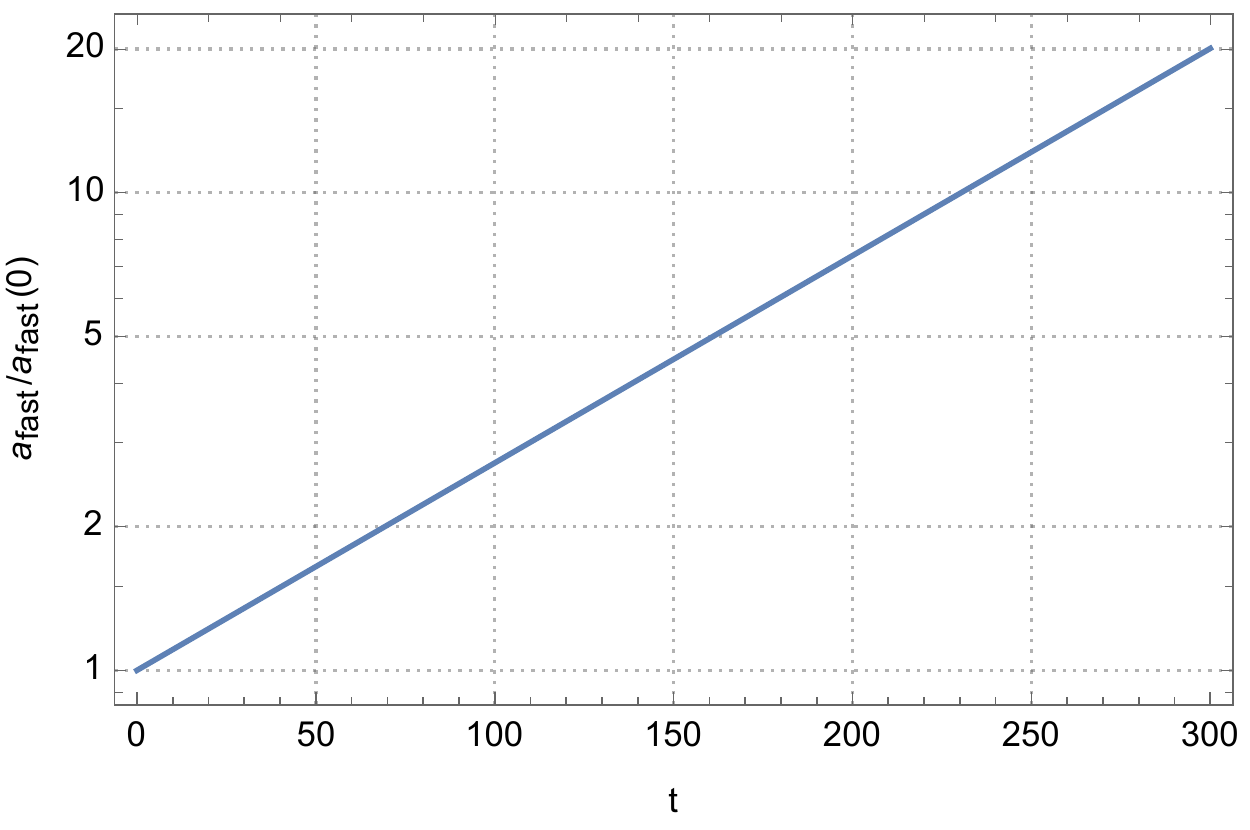}
\caption{Quantum pair production and oscillation 
in a sphere $H^{-3}_{\rm fast}$ are illustrated for pre-inflation,  
using $M=10^{-1}m_{\rm pl}$, $H_{\rm fast}(0)=10^{-2}m_{\rm pl}$, 
${\mathcal N}_{\rm pair}=10^2$, and $a_{\rm fast}(0)=1$ at initial time $t=0$. Oscillating Bogoliubov $|\beta|^2$ shows nontrivial particle productions. The time variation $\dot H_{\rm fast}\propto -(\rho_{_M}^{\rm fast} + p_{_M}^{\rm fast})$. $H_{\rm fast}$ and $\rho^{\rm fast}_{_\Lambda}$decrease very slowly. The scale factor $a_{\rm fast}$  increases exponentially. The Planck unit $m_{\rm pl}=(1/8\pi G)^{1/2}=1$ is adopted for presenting numerical results.
}\label{detailosci0}
\end{figure*}

\end{document}